\documentclass[twocolumn]{aastex7}

\usepackage{placeins}
\usepackage{hyperref}
\usepackage{upgreek}
\usepackage{color}
\usepackage{amsmath}
\usepackage{graphicx}
\shorttitle{KOINTREAU II: Two New Companions To Ophiuchus Stars}
\shortauthors{Walker et al.}

\begin{document}

\title{Keck Observations in the INfrared of Taurus and $\rho$ Oph Exoplanets And Ultracool dwarfs (KOINTREAU) II: Two Young Bound Companions to Ophiuchus Stars}

\author[0000-0001-7062-815X]{Samuel A. U. Walker}
\affiliation{Institute for Astronomy at the University of Hawai`i at M\={a}noa, 2680 Woodlawn Drive, Hawai`i, HI 96822, USA}
\email{swalk@hawaii.edu}

\author[0000-0003-2232-7664]{Michael C. Liu}
\affiliation{Institute for Astronomy at the University of Hawai`i at M\={a}noa, 2680 Woodlawn Drive, Hawai`i, HI 96822, USA}
\email{mliu@ifa.hawaii.edu}

\author[0000-0002-8895-4735]{Dimitri Mawet}
\affiliation{Department of Astronomy, California Institute of Technology, Pasadena, CA 91125, USA}
\affiliation{Jet Propulsion Laboratory, California Institute of Technology, 4800 Oak Grove Dr., Pasadena, CA 91109, USA}
\email{dmawet@astro.caltech.edu}

\author[0000-0001-6041-7092]{Mark W. Phillips}
\affiliation{Institute for Astronomy at the University of Hawai`i at M\={a}noa, 2680 Woodlawn Drive, Hawai`i, HI 96822, USA}
\affiliation{Institute for Astronomy, University of Edinburgh, Blackford Hill, Edinburgh, Scotland, EH9 3HJ}
\email{mark.phillips@roe.ac.uk}

\author[0000-0002-1838-4757]{Aniket Sanghi}
\altaffiliation{NSF Graduate Research Fellow}
\affiliation{Cahill Center for Astronomy and Astrophysics, California Institute of Technology, 1200 E. California Boulevard, MC 249-17, Pasadena, CA 91125, USA}
\email{asanghi@caltech.edu}

\author[0000-0003-1698-9696]{Bin B. Ren}
\affiliation{P.E.S., Observatoire de la Côte d'Azur, Nice 06304, France}
\email{bin.ren@oca.eu}

\author[0000-0002-6879-3030]{Taichi Uyama}
\affiliation{California State University Northridge, 8111 Nordhoff Street, Northridge, CA 91330-8268}
\email{taichi.uyama.astro@gmail.com}



\begin{abstract}
\noindent
We present the second set of discoveries from Keck Observations in the INfrared of Taurus and $\rho$ Oph Exoplanets And Ultracool dwarfs (KOINTREAU), an adaptive optics survey of young stars in the Taurus and $\rho$ Oph star-forming regions using Keck/NIRC2 in conjunction with the Keck infrared pyramid wavefront sensor.
We have discovered two faint comoving companions to young stars ISO-Oph~96 and 2MASS~J16262785$-$2625152. The companion to ISO-Oph~96, KOINTREAU\nobreakdash-\hspace{0pt}3b, is at a projected separation of 340 au (2.49$^{\prime\prime}$). Using our NIRC2 photometry and evolutionary models, and assuming that the companion has the same extinction as its host star, we infer that KOINTREAU\nobreakdash-\hspace{0pt}3b has a mass of $3.4\pm0.7$~M$_{\rm Jup}$. The companion to 2MASS~J16262785$-$2625152, KOINTREAU\nobreakdash-\hspace{0pt}4b, has a projected separation of 180 au (1.25$^{\prime\prime}$) and could have a mass of either $11.5^{+1.2}_{-1.6}$~M$_{\rm Jup}$ or $15.3^{+0.7}_{-0.8}$~M$_{\rm Jup}$, depending on whether the host star is a member of $\rho$ Oph or Upper Sco.
\end{abstract}



\section{Introduction} \label{sec:intro}

Exoplanet direct imaging is uniquely sensitive to massive companions ($\gtrsim$1~M$_{\rm Jup}$) on wide orbits ($\gtrsim$10 au), probing a regime of the broader exoplanet population that other detection and characterization methods struggle to reach \citep{bowler2016}.
Indeed, direct imaging has led to the discovery of companions across a wide range of orbital separations, from $<$10 au \citep[e.g., AF Lep b, $\beta$ Pic b, and 51 Eri b;][]{derosa2023, franson2023, mesa2023, lagrange2009, macintosh2015} to $>$1000 au \citep[e.g., Ross 458 c, GU Psc b, and COCONUTS-2b;][]{goldman2010, naud2014, zhang2021}.

Imaging extremely young exoplanets ($\lesssim$10 Myr) is of particular interest, not only because of the favorable planet/star contrast ratio at early times but also due to the ability of direct imaging to probe the relationship between forming planets and their natal protoplanetary disk \citep{keppler2018, haffert2019, benisty2021}. Furthermore, it is at these young ages that the differences in predicted physical properties between exoplanet formation scenarios are greatest \citep[e.g.,][]{spiegelandburrows2012}, and thus imaging and characterizing young exoplanets can grant key insights into their formation.

Young moving groups (YMGs; e.g., $\beta$ Pictoris, TW Hydrae) have been a popular target of exoplanet imaging surveys \citep[e.g.,][]{biller2013, galicher2016, nielsen2019, vigan2021, sanghi2024} due to their proximity and youth \citep[e.g., the $\beta$ Pic moving group, $\approx$35 pc and $\approx$25 Myr;][]{bpmgage}, and indeed these regions have produced a number of exoplanet discoveries \citep[e.g.,][]{marois2008, lagrange2009, bowler2015, macintosh2015, derosa2023, franson2023, mesa2023}. Star-forming regions such as Taurus and Ophiuchus are younger ($\approx$1-5 Myr) and less thoroughly characterized than the YMGs, making them prime targets for imaging surveys.

The Ophiuchus star-forming region ($\sim$2-5 Myr, 140 pc), also known as $\rho$ Oph, has been previously surveyed on a range of scales. On wide angular scales, there have been searches for member stars using data from missions such as \textit{Gaia} \citep{gaiadr3}, the 2 Micron All Sky Survey \citep[2MASS;][]{cutri2003, skrutskie2006} and the \textit{Spitzer Space Telescope} \citep{werner04, irac}, and complemented by spectroscopic follow-up \citep[e.g.,][]{padgett2008, shirono2011, esplin2020, grasser2021, miret-roig2022}. There have also been a number of ground-based wide-field imaging surveys of Ophiuchus \citep[e.g.,][]{oliveira2008, haisch2010, geers2011, barsony2012, muzic2012, chiang2015, allers2020}. Key findings of these surveys include the discovery that there is a significantly greater number of free-floating planet (FFP) candidates in Ophiuchus than predicted by core-collapse formation models \citep{miret-roig2022}, indicating that system ejection might be a significant contributor to the formation of FFPs, and the determinations by \citet{muzic2012} and \citet{esplin2020} that $\sim$40-50\% of Ophiuchus stars host disks.

On smaller angular scales, there have been optical/near-infrared imaging surveys using large-aperture telescopes to observe systems on an individual basis, either with adaptive optics \citep[AO;][]{correia2006, close2007, duchene2007, cieza2010} or employing AO combined with sparse aperture masking to aid in the detection of the very closest companions \citep{cheetham2015}. AO imaging surveys attain deeper contrasts than non-AO surveys, which makes them a particularly useful tool in constraining the occurrence rates of substellar objects on wide orbits. For example, \citet{cheetham2015} ascertained that $7^{+8}_{-5}$\% of Ophiuchus stars have brown dwarf companions between 1.3 and 780 au. Such work can then be used to constrain substellar formation models.

One quandary that surveyors of Ophiuchus face is exactly how to define the star-forming region's extent, as Ophiuchus is co-located on the sky with the eastern edge of the Upper Scorpius association, an older \citep[10 Myr;][]{pecaut2016} subgroup of the Scorpius-Centaurus OB association to which Ophiuchus also belongs. Age is a key variable in substellar evolutionary models, particularly at these young ages, and thus whether an individual star belongs to Ophiuchus or Upper Sco can significantly impact the derived physical properties of substellar companions. The distinction between these two regions is typically drawn using \textit{Gaia} kinematics. While being satisfactory on a population level, this approach may not be equally reliable for all stars in the region. For example, $\rho$ Oph B, despite being an accepted Ophiuchus member and a 3$^{\prime\prime}$ companion of the star $\rho$ Oph, does not satisfy the proper motion criteria used by \citet{luhman2020}.


To date, there have been four directly-imaged companions discovered in Ophiuchus with masses below or within 1$\sigma$ of the canonical 13~M$_{\rm Jup}$ boundary between an exoplanet and brown dwarf (hereinafter planetary-mass companions, or PMCs): SR12 C \citep[1100 au projected separation, 13.6~$\pm$~7~M$_{\rm Jup}$,][]{kuzuhara2011}, ROXs 42B b \citep[150 au, 10~$\pm$~4~M$_{\rm Jup}$,][]{currie2014, kraus2014}, CFHTWIR-Oph~98~B \citep[200 au, 7.8~$\pm$~0.8~M$_{\rm Jup}$,][]{fontanive2020} and ROXs 12 b \citep[210 au, 16~$\pm$~4~M$_{\rm Jup}$,][]{kraus2014}.

This publication is the second in a series of papers based on Keck Observations in the INfrared of Taurus and $\rho$ Oph Exoplanets And Ultracool dwarfs (KOINTREAU), an adaptive optics survey of young stars in Taurus and $\rho$ Oph using the Keck infrared pyramid wavefront sensor (PyWFS). The first publication \citep[][hereinafter Paper I]{kointreauI}, detailed the methodology of our survey and presented the discovery of two faint Taurus companions: KOINTREAU\nobreakdash-\hspace{0pt}1b, an M9 \textsc{vl-g} companion with an estimated mass of $10.6^{+2.5}_{-2.3}$~M$_{\rm Jup}$, and KOINTREAU\nobreakdash-\hspace{0pt}2b, an M4.5 young star occulted by an edge-on disk. 

\begin{deluxetable}{lcc}
\centering
\tabletypesize{\small}
\tablecaption{ISO-Oph 96 system properties\label{tab:isooph96}}
\tablehead{\colhead{Property} & \colhead{Value}& \colhead{Ref.}}
\startdata
\multicolumn{3}{c}{ISO-Oph 96}\\
\hline
RA (J2000)&16 27 04.52& (2)\\
Dec. (J2000)&$-$24 42 59.67& (2)\\
$\mu_{\rm RA}$ (mas yr$^{-1}$) & $-6.5\pm0.1$ & (2)\\
$\mu_{\rm Dec.}$ (mas yr$^{-1}$) &$-24.97\pm0.09$& (2)\\
$\pi$ (mas) & $7.26\pm0.08$ & (2)\\
Distance (pc) & $138 \pm 2$ & (2)\\
Gaia DR3 RUWE & 0.964 & (2)\\
Spectral type & K6-M1 & (3)\\
$T_{\rm eff}$ (K) & $3920\pm70$ & (4)\\
log g (cm s$^{-2}$) & $4.4\pm0.1$ & (4)\\
{[Fe/H]} (dex) & $-0.05\pm0.01$ & (4)\\
$v\sin{i}$ (km s$^{-1}$) & $16.9\pm0.6$ & (5)\\
$A_K$ (mag)& $0.73$ & (6)\\
\textit{Gaia} $G$ (mag) & $16.586\pm0.005$ & (2)\\	
$J_{\rm MKO}$ (mag) & $10.65\pm0.03$ & (3,4)\tablenotemark{a}\\
$H_{\rm MKO}$ (mag) & $9.43\pm0.03$ & (3,4)\tablenotemark{a}\\
$K_{\rm MKO}$ (mag) & $9.78\pm0.03$ & (7,8)\tablenotemark{a}\\
\hline
\multicolumn{3}{c}{KOINTREAU\nobreakdash-\hspace{0pt}3b}\\
\hline
Projected separation (au) & $342\pm4$ & (1)\\
$K_{\rm MKO}$ (mag)& $17.9\pm0.1$ & (1)\\
Estimated spectral type & $>$L0 & (1)\\
Mass (M$_{\rm Jup}$) &$3.4\pm0.7$ & (1)\\
\enddata
\tablenotetext{a}{Transformed from 2MASS magnitudes using relations given in  \citet{skrutskie2006}.}
\tablerefs{
(1) this work;
(2) \citet{gaiadr3};
(3) \citet{erickson2011};
(4) \citet{jonsson2020};
(5) \citet{hourihane2023};
(6) \citet{luhman2020};
(7) \citet{cutri2003};
(8) \citet{skrutskie2006}
}
\end{deluxetable}

\begin{deluxetable}{lcc}
\centering
\tabletypesize{\small}
\tablecaption{2MASS J1626$-$2625 system properties\label{tab:2mj1626}}
\tablehead{\colhead{Property} & \colhead{Value}& \colhead{Ref.}}
\startdata
\multicolumn{3}{c}{2MASS~J16262785$-$2625152}\\
\hline
RA (J2000)&16 26 27.86& (2)\\
Dec. (J2000)&$-$26 25 15.18& (2)\\
$\mu_{\rm RA}$ (mas yr$^{-1}$) & $-7.14\pm0.06$ & (2)\\
$\mu_{\rm Dec.}$ (mas yr$^{-1}$) & $-26.18\pm0.05$ & (2)\\
$\pi$ (mas) & $7.05\pm0.05$ & (2)\\
Distance (pc) & $142\pm1$ & (2)\\
Gaia DR3 RUWE & 2.360 & (2)\\
Spectral type & M4.5 & (3)\\
$A_K$ (mag)& $0.06$ & (3)\\
\textit{Gaia} $G$ (mag) & $14.057\pm0.006$ & (2)\\	
$J_{\rm MKO}$ (mag)& $10.78\pm0.03$ & (4,5)\tablenotemark{a}\\
$H_{\rm MKO}$ (mag)& $10.18\pm0.03$ & (4,5)\tablenotemark{a}\\
$K_{\rm MKO}$ (mag)& $9.82\pm0.03$ & (4,5)\tablenotemark{a}\\
\hline
\multicolumn{3}{c}{KOINTREAU\nobreakdash-\hspace{0pt}4b}\\
\hline
Projected separation (au) & $182\pm2$ & (1)\\
$J_{\rm MKO}$ (mag) & $16.67\pm0.08$ & (1)\\
$H_{\rm MKO}$ (mag) & $16.28\pm0.06$ & (1)\\
$K_{\rm MKO}$ (mag) & $15.6\pm0.1$ & (1)\\
Estimated spectral type & L0 $\pm$ 2 & (1)\\
Mass if Oph member (M$_{\rm Jup}$) & $11.5^{+1.2}_{-1.6}$ & (1)\\
Mass if USco member (M$_{\rm Jup}$) & $15.3^{+0.7}_{-0.8}$ & (1)\\
\enddata
\tablenotetext{a}{Transformed from 2MASS magnitudes using relations given in  \citet{skrutskie2006}.}
\tablerefs{
(1) this work;
(2) \citet{gaiadr3};
(3) \citet{luhman2020};
(4) \citet{cutri2003};
(5) \citet{skrutskie2006}
}
\end{deluxetable}

In this paper, we report our first Ophiuchus discoveries from KOINTREAU, two faint companions around ISO-Oph~96 and 2MASS J16262785$-$2625152, and the subsequent confirmation of their status as bound companions. We then characterize these companions using their Keck/NIRC2 photometry. Our data show that KOINTREAU\nobreakdash-\hspace{0pt}3b, our imaged companion to ISO-Oph~96, may be the lowest-mass directly imaged companion in Ophiuchus assuming the companion experiences the same extinction as its host star. The companion to 2MASS~J16262785$-$2625152, KOINTREAU\nobreakdash-\hspace{0pt}4b, has a mass which straddles the boundary between exoplanet and brown dwarf.


\section{Target Stars}\label{sec:targs}

\subsection{ISO-Oph~96}

ISO-Oph~96 has been associated with the $\rho$ Ophiuchi molecular cloud since early near-infrared (NIR) imaging surveys of the region \citep{greene1992}. With the advent of \textit{Gaia}, this association has been kinematically verified by a number of studies \citep{esplin2020, gutierrezalbarran2020, jackson2022, miret-roig2022}. \citet{bontemps2001} used the Infrared Space Observatory camera (ISOCAM) to characterize ISO-Oph~96 as a Class III young stellar object (YSO), although we note that \citet{esplin2020} finds no evidence for a disk based on a lack of excess emission in IRAC or Wide-field Infrared Survey Explorer (WISE) data of this source. \citet{esplin2020} reports $A_K = 0.73$ mag based on its 2MASS $J-H$ color.
An optical spectrum by \citet{erickson2011} shows ISO-Oph~96 to have a late-K spectral type (K6-M1, which they average to adopt a spectral type of K8), and to exhibit lithium absorption and H$\alpha$ emission, both of which corroborate its youth \citep[see also][]{gutierrezalbarran2020}.  
\citet{manara2015} report a mass and effective temperature for ISO-Oph~96 of 0.35 M$_\odot$ and 3311K (M3.5), conflicting with the \citet{erickson2011} spectral type, but this is based solely on using photometry \citep{natta2006}. We adopt the \citet{erickson2011} spectroscopically-derived value following \citet{esplin2020}.

As for ISO-Oph~96's age, \citet{kerr2021} identify it as a bona fide $\rho$ Oph member using \textit{Gaia} data, and use isochrone fitting to age the $\rho$ Oph cluster to $5.7\pm0.4$ Myr. However, \citet{kerr2021} also estimates ages for individual stars by comparing with synthetic stellar populations, and this gives an age for the star of $11.0^{+2.6}_{-3.1}$ Myr.
\citet{grasser2021} finds the star to belong to their main Ophiuchus population, to which they assign an age of 0.3-6 Myr.
\citet{esplin2020} attribute ISO-Oph~96 to a higher-extinction ($A_J\geq1.5$) subset of the $\rho$ Oph subcluster L1688, which they age-date to $2\pm1$ Myr by assuming an age of 10 Myr for Upper Sco, measuring the difference in average $K$-band magnitude between the high-extinction L1688 members and their Upper Sco group, and converting this to a difference in age using the BHAC15 evolutionary models \citep{bhac2015}. Finally, the BANYAN~$\Sigma$ young stellar association membership code \citep{banyansigma} identifies ISO-Oph~96 as being a member of Upper Sco (which they age to $10\pm3$ Myr) with 87.8\% confidence, with an 11.8\% chance of it being a member of $\rho$ Oph (with an adopted age of $<2$ Myr).

To summarize, while the literature favors adopting ISO-Oph~96 as a member of Ophiuchus, there is still some uncertainty as to its age, with estimates ranging from $2\pm1$ Myr to $11.0^{+2.6}_{-3.1}$ Myr. Key properties of the ISO-Oph~96 system are reported in Table~\ref{tab:isooph96}.

\subsection{2M~1626\nobreakdash$-$\hspace{0pt}2625}

Our other target, 2MASS J16262785$-$2625152 (hereinafter 2M~1626\nobreakdash$-$\hspace{0pt}2625), is less well-studied than ISO-Oph~96, and as such its membership is less certain.

\citet{esplin2020} find 2M~1626\nobreakdash$-$\hspace{0pt}2625 to be in accord with other Ophiuchus sources both positionally and kinematically. They associate the source with an ``off-cloud'' subgroup which they age-date to $5\pm1$ Myr in the same manner as described above for L1688. Its off-cloud nature leads later work to associate 2M~1626\nobreakdash$-$\hspace{0pt}2625 with Upper Sco rather than Ophiuchus \citep{luhman2022}. They measured an optical spectral type of M4.5 and calculated $A_K=0.06$ mag by comparing this spectral type with its $J-H$ color. They also find no evidence for infrared excess indicative of a disk, although they only report detections in WISE W2 and W3 bands. Its optical spectrum also exhibits a lithium equivalent width of 500 m\AA, indicative of youth \citep{luhman2020}. In slight tension with \citet{esplin2020}, \citet{miret-roig2022} identifies 2M~1626\nobreakdash$-$\hspace{0pt}2625 as belonging to $\sigma$ Sco, a subgroup of Upper Sco that they find to have a dynamical traceback age of $2.1\pm0.7$ Myr, although it should be noted that this difference in ages is likely to arise from the differing methodology than the different group clusterings used in the two works. For example, \citeauthor{miret-roig2022} find $\rho$ Oph itself to have a dynamical traceback age consistent with zero. \citet{grasser2021} finds this star to belong to their main Ophiuchus population, as they do with ISO-Oph~96.
Complicating the picture further, \citet{kerr2021} assign 2M~1626\nobreakdash$-$\hspace{0pt}2625 to Upper Sco, which they give an isochrone-based age of $11.3\pm0.3$ Myr. However, the \citet{kerr2021} individual age estimate for 2M~1626\nobreakdash$-$\hspace{0pt}2625 gives $3.2^{+1.5}_{-0.6}$ Myr, which is more in line with their age estimate for Ophiuchus.
BANYAN~$\Sigma$ identifies 2M~1626\nobreakdash$-$\hspace{0pt}2625 as a member of Upper Sco with 96.1\% confidence.

In summary, there is disagreement in the literature about whether 2M~1626\nobreakdash$-$\hspace{0pt}2625 is a member of Ophiuchus or Upper Sco, which leads to disagreement as to its age, with cluster age estimates ranging from $5\pm1$ Myr to $11.3\pm0.3$ Myr. This system's properties are reported in Table \ref{tab:2mj1626}.

\begin{figure*}[t]
         \centering
         \includegraphics[width=0.49\linewidth]{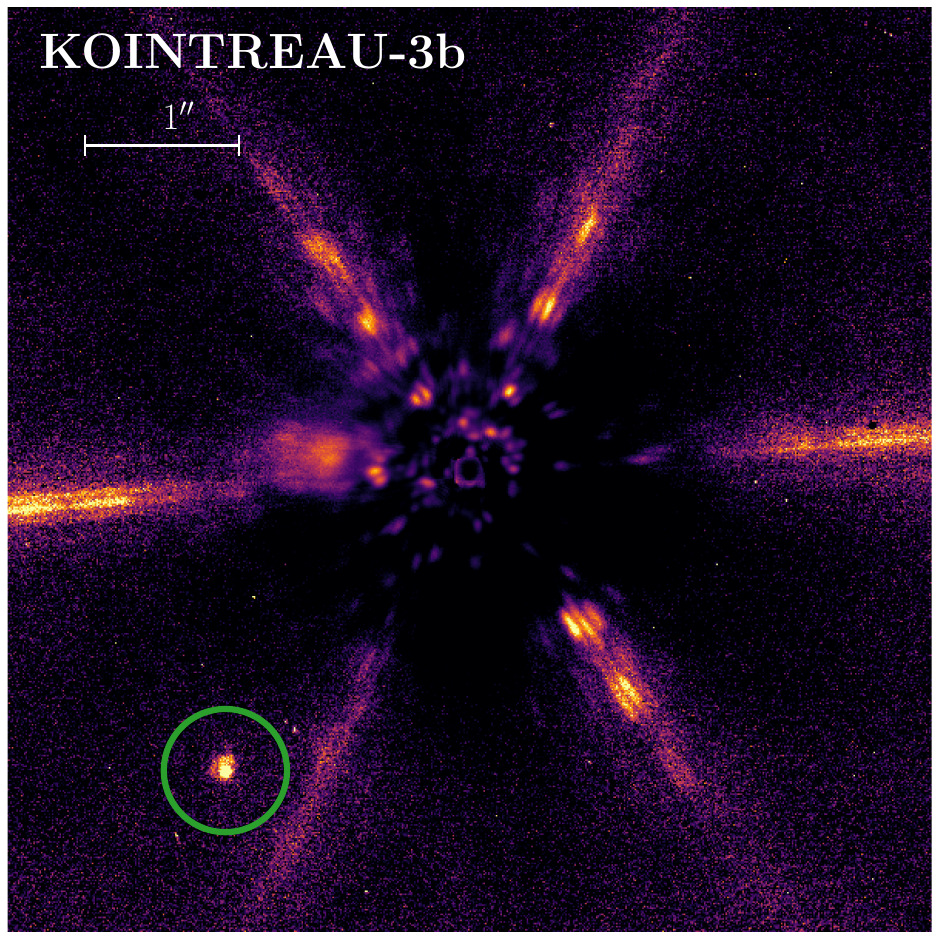}
         \includegraphics[width=0.49\linewidth]{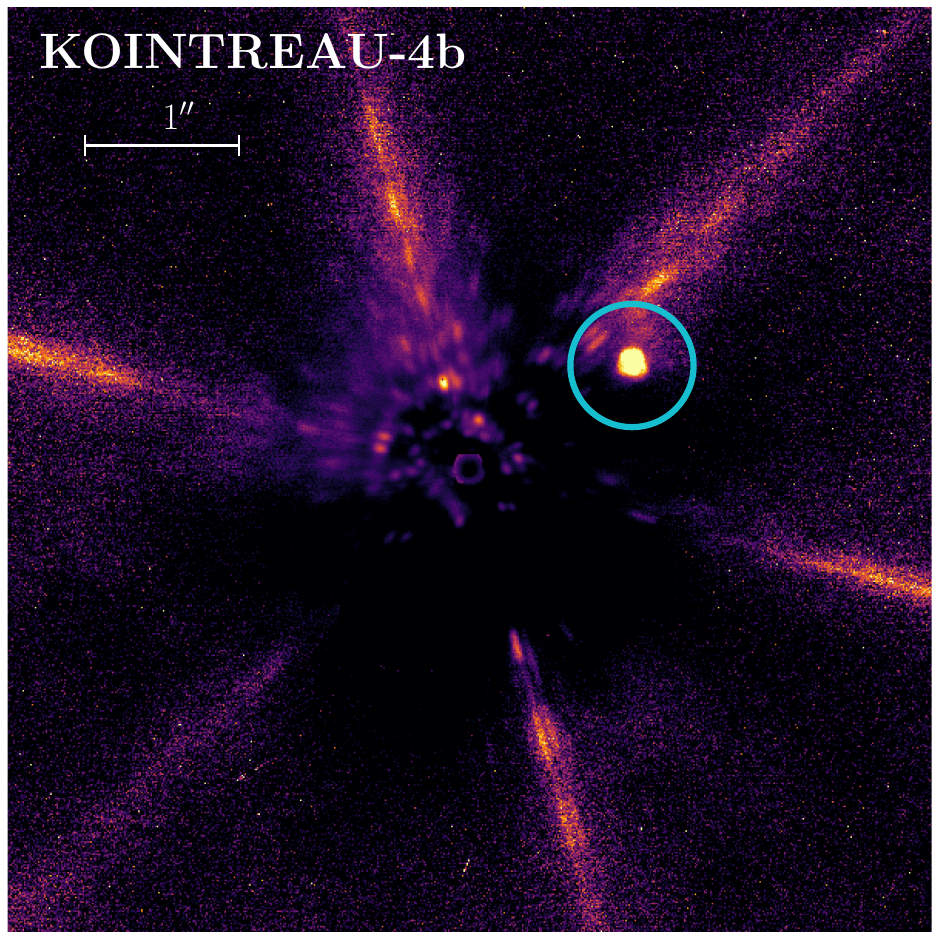}
     \caption{6$^{\prime\prime}$ cutouts from our Keck/NIRC2 signal-to-noise maps from our deep-exposure $K$-band image stacks of each candidate companion, oriented North-up East-left. These maps are computed by subtracting the median and normalizing by the robust standard deviation in successive annuli around the host star. The candidate companions are circled.}
    \label{fig:nirc2}
\end{figure*}

\section{Keck/NIRC2 Imaging} \label{sec:imaging}

We imaged our two target stars with Keck/NIRC2 and adaptive optics, using the infrared pyramid wavefront sensor \citep[PyWFS;][]{pywfsjatis} with a natural guide star (NGS) in 2022 and the Shack-Hartmann wavefront sensor \citep[SHWFS;][]{wizinowich2000} in both NGS and laser guide star \citep[LGS;][]{wizinowich2006, vandam2006} modes in 2024 and 2025. Example detections are shown in Figure~\ref{fig:nirc2}, and our observations are summarized in Table~\ref{tab:nirc2}. Our observing sequence consisted of a set of short exposures, used to register the unsaturated position of the host star, then a set of longer exposures to detect any faint companions. We took data in the Mauna Kea Observatories \citep[MKO;][]{simons2002,tokunaga2002} $K$-band filter at each epoch, and also obtained MKO $JH$ data during our April 2024 observing run for 2M~1626\nobreakdash$-$\hspace{0pt}2625. For further details on our observing and data reduction procedure, see Paper I. Our observations resulted in the detection of one companion of interest around each of ISO-Oph~96 and 2M~1626\nobreakdash$-$\hspace{0pt}2625, which we refer to henceforth as KOINTREAU\nobreakdash-\hspace{0pt}3b and KOINTREAU\nobreakdash-\hspace{0pt}4b, respectively.

We centroided each host star in their short-exposure images, and each companion in their long-exposure images, took the mean and standard deviation of each set of measurements, and used these to compute the relative astrometry between each companion and its host star per epoch. We corrected for NIRC2's optical distortion using a combination of the \citet{service2016} distortion correction and an additional position angle offset of 0.118$^{\circ}$ derived from our analysis of images of M92 (see Paper I).

We also measured the contrast of each candidate companion relative to its host star following the procedure described in Paper I (Table \ref{tab:nirc2}). Briefly, we performed aperture photometry on the host star in the short-exposure images and the companion in the long-exposure images, then took the ratio of the counts per second for the two objects to obtain a contrast value for each object and epoch. We used variable aperture sizes so as to maximize the signal-to-noise of the resulting contrast measurement for each epoch. For the deep-exposure images, we also median-subtracted a radial profile from successive annuli around the host star to remove the light from the host star before measuring the brightness of the companion.

KOINTREAU\nobreakdash-\hspace{0pt}3b is sufficiently separated from its host star that the above approach proved satisfactory. However, KOINTREAU\nobreakdash-\hspace{0pt}4b is only $\approx1.25\arcsec$ from its host star, and additionally its position angle relative to its host star meant that for multiple observations it landed on or near one of the diffraction spikes from the host star (e.g. Figure~\ref{fig:nirc2}), making the described process insufficient for removing the starlight from our flux measurements of this companion. To counter this, we took advantage of the symmetry of the NIRC2 point spread function (PSF), and measured the flux in identically-sized apertures at $\pm60^{\circ}$ and $\pm120^{\circ}$ from KOINTREAU\nobreakdash-\hspace{0pt}4b's position angle. We took the average of these measurements as the estimated background for the companion, and subtracted it accordingly from the measured flux of the companion before computing the contrast.

We adopt the average of the 2022 and 2024 $K$-band contrasts for KOINTREAU\nobreakdash-\hspace{0pt}3b and the 2024 $K$-band contrast for KOINTREAU\nobreakdash-\hspace{0pt}4b in our subsequent calculations (although we note that all $K$-band contrasts are consistent for both sources).
We then transformed the 2MASS $JHK_S$ magnitudes of the host stars to MKO \citep{skrutskie2006} and added these to our measured contrasts to determine the companions' magnitudes (Tables~\ref{tab:isooph96} and~\ref{tab:2mj1626}).

\begin{rotatetable*}
\begin{deluxetable*}{lcccccccccc}
\centering
\tabletypesize{\footnotesize}
\tablecaption{NIRC2 AO Imaging Summary\label{tab:nirc2}}
\tablehead{\colhead{Target} & \colhead{Date} & \colhead{Filter} & Camera & \colhead{Shallow} & \colhead{Deep} & \colhead{AO mode/} & \colhead{Separation} & \colhead{Position Angle} & \colhead{Contrast} & \colhead{FWHM}\\
\colhead{} & \colhead{(UT)} & \colhead{(MKO)} & & \colhead{exp. time\tablenotemark{a}} & \colhead{ exp. time\tablenotemark{a}} & \colhead{WFS\tablenotemark{b}} & \colhead{(mas)} & \colhead{(deg)} & \colhead{($\Delta$mag)} & \colhead{(mas)\tablenotemark{c}}}
\startdata
KOINTREAU\nobreakdash-\hspace{0pt}3b& 2022-06-10 & $K$ & Narrow & $2\times10\times1$ & $3\times1\times60$ & NGS/Py & $2488\pm3$ & $140.54\pm0.09$ &$8.2\pm0.2$&$44\pm6$\\
& 2024-04-28 & $K$ & Narrow & $5\times10\times1$ & $3\times1\times60$ & LGS/SH & $2487\pm13$ & $140.36\pm0.3$ &$8.0\pm0.2$&$41\pm3$\\
\hline
KOINTREAU\nobreakdash-\hspace{0pt}4b& 2022-06-12 & $K$ & Narrow & $3\times10\times1$ & $3\times1\times60$ & NGS/Py & $1248\pm2$ & $302.27\pm0.01$ &$5.9\pm0.1$&$37\pm2$\\
& 2024-04-23 & $K$ & Narrow & $4\times50\times0.17$ & $3\times1\times60$ & LGS/SH & $1241\pm6$ & $302.1\pm0.4$ &$5.83\pm0.09$&$29\pm1$\\
& 2024-04-28 & $J$ & Narrow & $3\times5\times2$ & $3\times1\times60$ & LGS/SH & $1253\pm3$ & $302.2\pm0.2$ & $6.4\pm0.1$ & $53\pm3$\\
& 2024-04-28 & $H$ & Narrow & $3\times10\times1$ & $3\times1\times60$ & LGS/SH & $1278\pm6$ & $302.8\pm0.2$ & $6.02\pm0.09$ & $41\pm1$\\
& 2025-07-01 & $K$ & Narrow & $10\times50\times0.17$ & $5\times1\times60$ & NGS/SH & $1257\pm7$ & $302.6\pm0.4$ &$5.8\pm0.2$&$31\pm3$\\
\enddata
\tablenotetext{a}{Number of exposures $\times$ Number of coadds $\times$ Integration time per coadd (s)}
\tablenotetext{b}{Natural guide star (NGS) or laser guide star (LGS) mode using the Pyramid wavefront sensor (Py) or the Shack-Hartmann wavefront sensor (SH)}
\tablenotetext{c}{Calculated from the radial profile of the host star in the shallow-exposure stack.}
\end{deluxetable*}
\end{rotatetable*}

\begin{figure*}
    \centering
    \includegraphics[width=.9\linewidth]{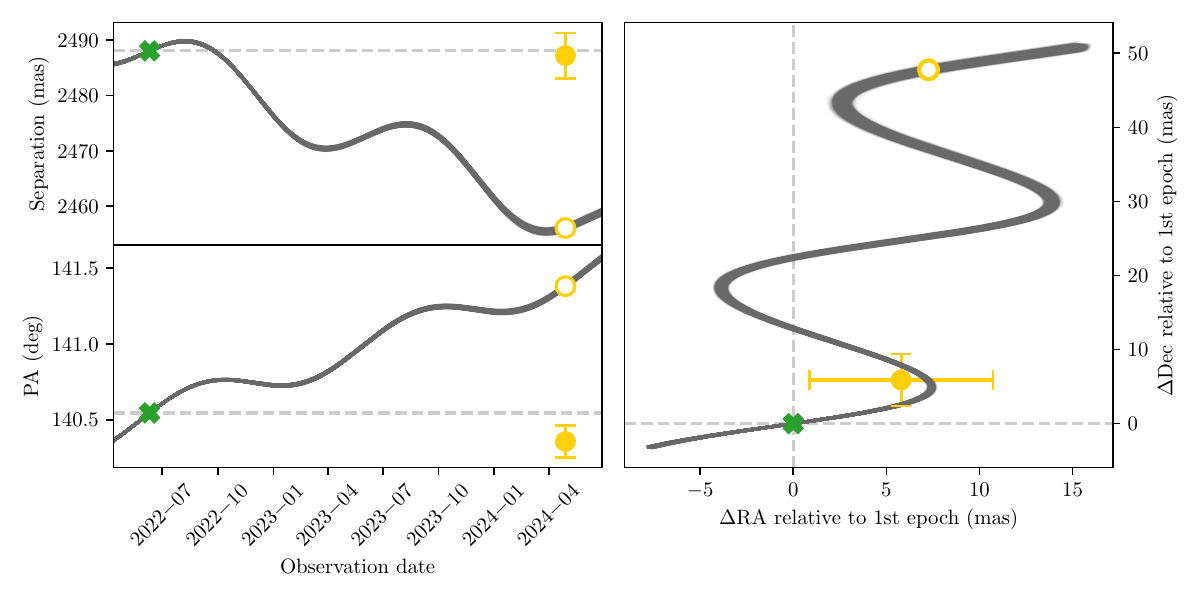}
    \caption{Relative astrometry of KOINTREAU\nobreakdash-\hspace{0pt}3b, the companion to ISO-Oph~96, demonstrating that it is physically associated with ISO-Oph~96 and not a background object. The dark gray curve shows 1000 draws from a model of the predicted movement of a background object relative to the companion's first epoch detection (the cross and gray dashed lines in each panel).
    The hollow markers indicate where the companion would appear if it were a background object at each epoch, and the filled markers indicate the measured position of the companion in the corresponding epoch. Note that measurements are presented relative to the first epoch, hence first-epoch measurement errors are incorporated into the displayed errorbars for subsequent epochs.}
    \label{fig:isooph96_astrometry}
\end{figure*}
\begin{figure*}
    \centering
    \includegraphics[width=.9\linewidth]{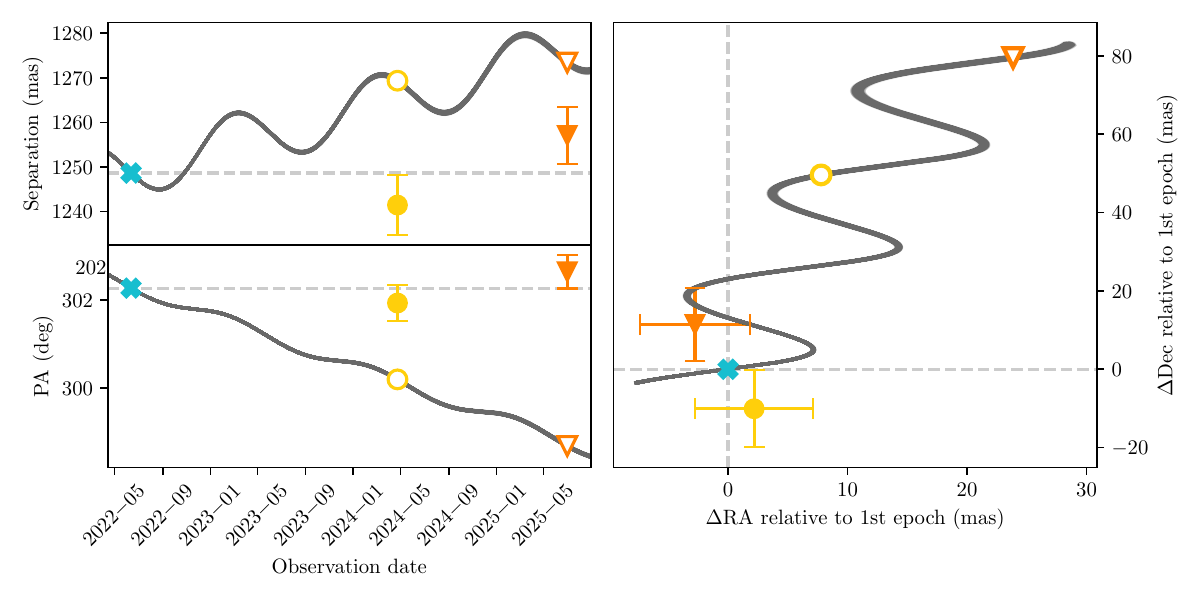}
    \caption{Relative astrometry for KOINTREAU\nobreakdash-\hspace{0pt}4b, our observed companion to 2M~1626\nobreakdash$-$\hspace{0pt}2625, demonstrating that it is physically associated with 2M~1626\nobreakdash$-$\hspace{0pt}2625 and not a background object. See Figure~\ref{fig:isooph96_astrometry} caption for full description.}
    \label{fig:2mj1626_astrometry}
\end{figure*}

\section{Results}

\subsection{Common Proper Motion}

We combined our NIRC2 astrometry of the companions with proper motion and parallax data for their host stars from \textit{Gaia} DR3 to ascertain if the candidates are comoving with their central stars (Figures~\ref{fig:isooph96_astrometry} and~\ref{fig:2mj1626_astrometry}).
The measured sky motion of each candidate companion relative to its host over time is consistent with zero. We confirm this by computing $\chi^2_\nu$ for the background object and comoving companion hypotheses and comparing these via the Bayes factor following \citet{bowler2013}. For KOINTREAU\nobreakdash-\hspace{0pt}3b, we obtain $\chi^2_{\nu, {\rm background}}=464$ and $\chi^2_{\nu, {\rm comoving}}=3.2$ (1 degree of freedom), which gives a log(Bayes factor) of 100.0, indicating a decisive preference for the comoving companion hypothesis using Jeffreys' scale \citep{jeffreys1961theory}. The same analysis for KOINTREAU\nobreakdash-\hspace{0pt}4b gives $\chi^2_{\nu, {\rm background}}=81.5$ and $\chi^2_{\nu, {\rm comoving}}=2.07$ (2 degrees of freedom), decisively confirming the comoving companion hypothesis with a log(Bayes factor) of 34.5.

To better understand the likelihood of these discoveries, we computed the likelihood that we would have found a background star that would appear like a substellar comoving companion at any point during our survey of Ophiuchus so far. We note that this is not the same as testing whether we would have found a background star that would have been confused for either of the companions we have discovered, for which the odds are significantly lower. For this test, we considered a companion of interest to have a separation to its host star $<5\arcsec$, and to have a mass $<$60~M$_{\rm Jup}$ (i.e. substellar), which approximately corresponds to an upper limit on apparent $K$-band brightness of 11.75 mag given the distance and age of Ophiuchus. We adopted a lower magnitude limit of $K=21$ based on a median limiting contrast of $\Delta K\approx11$ as computed from our survey data and assuming a typical host star brightness of $K=10$. We also adopted a fiducial error in measured companion proper motion of 2.5 mas/yr in RA and Dec. based on our typical measurement errors and time baseline (Figures \ref{fig:isooph96_astrometry} and \ref{fig:2mj1626_astrometry}).

We simulated a set of background stars over a sky area of 1 square degree in the direction of Ophiuchus using the Besan\c{c}on stellar population model \citep{besancon}. Then, we computed the number of background stars in this sample with $11.75<K<21$ that had proper motions in RA and Dec. within 2.5 mas/yr of the proper motions of each of the stars we have observed in Ophiuchus. We summed this value over all our observed target stars and scaled it to the sky-projected area of a circle of radius $5\arcsec$ per star. This gave a 2\% chance that we might find one or more background objects that might be confused with a bound substellar companion across our entire survey area to date (or a 0.018\% chance that we might find such an object around a given star in our sample). Further, given the 93 sources fainter than $K=11.75$ that were identified across Ophiuchus by \citet{esplin2020}, the likelihood of a chance alignment of another Ophiuchus source of interest within $5\arcsec$ of a given star in the sample is 0.013\% (a 1.5\% chance to happen at least once across our observed targets). The low likelihood that we would have erroneously identified a background star or another Ophiuchus member as a bound substellar companion to one of our survey stars increases our confidence that KOINTREAU\nobreakdash-\hspace{0pt}3b and KOINTREAU\nobreakdash-\hspace{0pt}4b are bona fide bound companions to their respective host stars.

We note that, although the change in position of each candidate is consistent with zero, our errorbars do not rule out super-escape velocities for these companions. KOINTREAU-3b's relative proper motion corresponds to a sky-projected speed relative to its host star of $2.6\pm1.5$ kms$^{-1}$, compared to an escape velocity given its projected separation and the mass of its host star of $1.7\pm0.2$ kms$^{-1}$. Likewise, KOINTREAU-4b's proper motion corresponds to a speed of $2.6\pm2$ kms$^{-1}$, compared to an escape velocity of $2.1\pm0.3$ kms$^{-1}$. This does not affect our conclusion that both objects are bound.

\begin{figure*}[t]
    \centering
    \includegraphics[width=.455\linewidth]{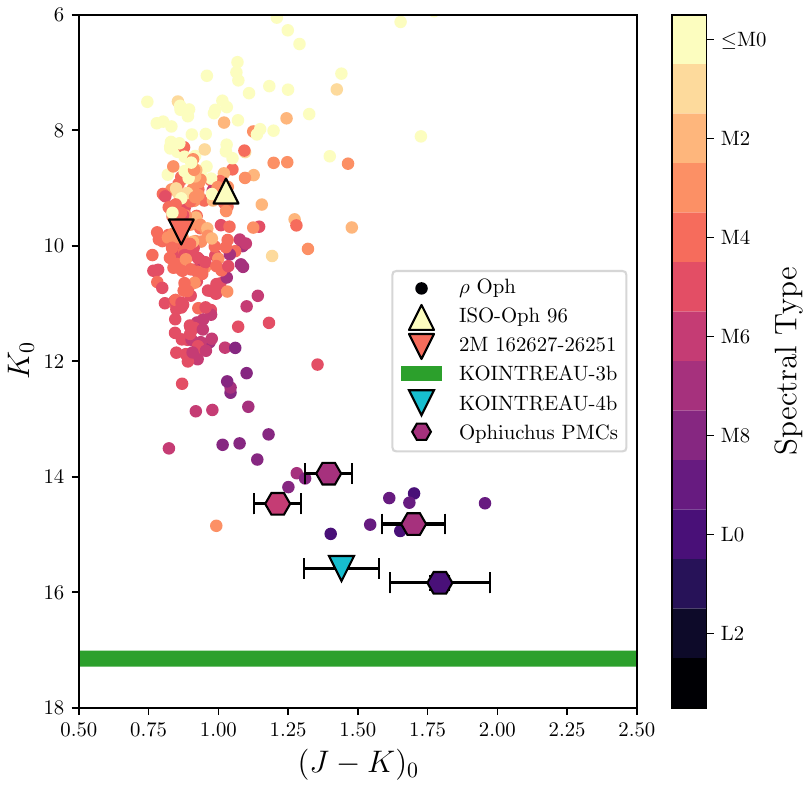}
    \includegraphics[width=.455\linewidth]{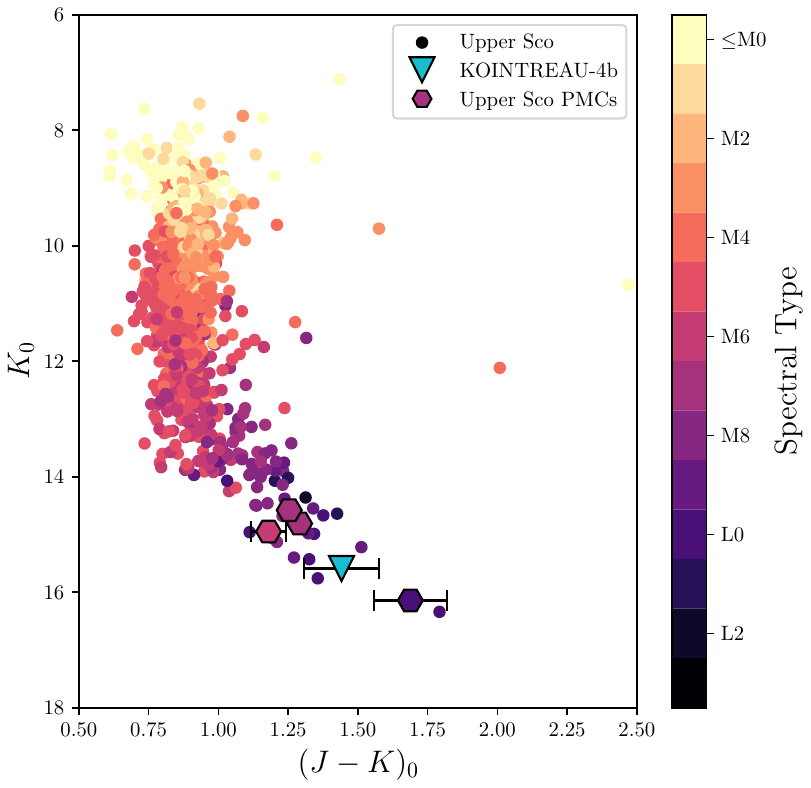}
    \caption{$J-K$ colour vs. extinction-corrected apparent $K$-band magnitude for our two candidate companions and their host stars, compared with Ophiuchus sources with spectral types K4-L3 from \citet{esplin2020} and known Ophiuchus planetary-mass companions (PMCs; left), and K4-L3 Upper Sco sources \citep{luhman2020} and PMCs (right). For KOINTREAU\nobreakdash-\hspace{0pt}3b we have no color data and thus indicate the dereddened $K$-band magnitude using a $\pm1\sigma$-tall band. 2MASS literature photometry has been transformed to MKO per \citet{skrutskie2006}. Ophiuchus PMCs are, from brightest to faintest: ROXs 12 b, \citeauthor{kraus2014}, \citeyear{kraus2014}; SR12 C, \citeauthor{kuzuhara2011}, \citeyear{kuzuhara2011}; ROXs 42B b, \citeauthor{kraus2014}, \citeyear{kraus2014}; CFHTWIR-Oph 98 B, \citeauthor{fontanive2020}, \citeyear{fontanive2020}. Upper Sco PMCs are, from brightest to faintest: USco 1621 B, \citeauthor{chinchilla2020}, \citeyear{chinchilla2020}; USco 1556 B, \citeauthor{chinchilla2020}, \citeyear{chinchilla2020}; GSC 06214-00210B, \citeauthor{ireland2011}, \citeyear{ireland2011}; 1RXS J160929.1-210524B, \citeauthor{lafreniere2008}, \citeyear{lafreniere2008}.}
    \label{fig:cmd_byspt}
\end{figure*}
\begin{figure*}[t]
    \centering
    \includegraphics[width=.455\linewidth]{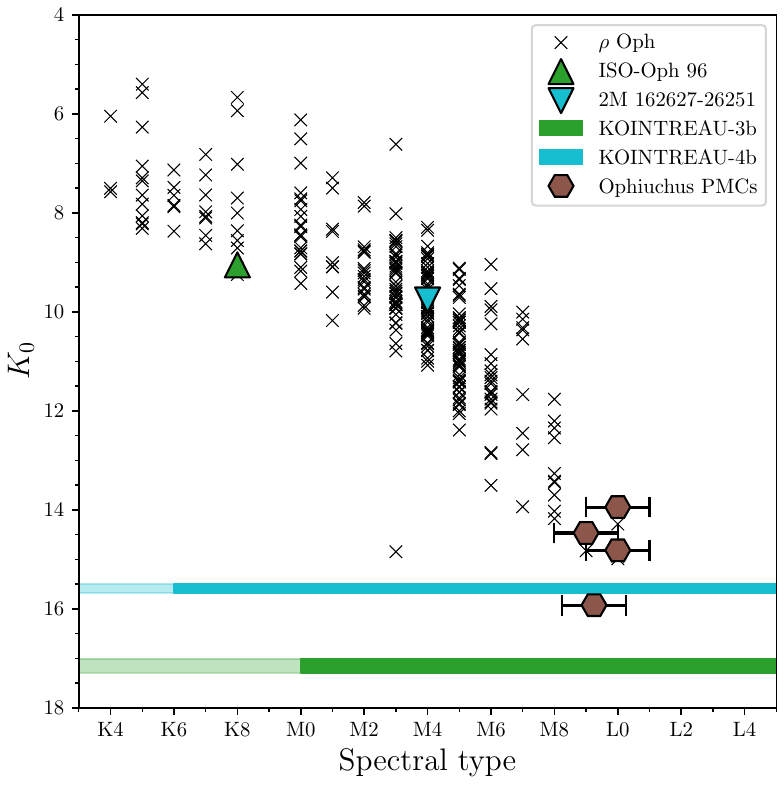}
    \includegraphics[width=.455\linewidth]{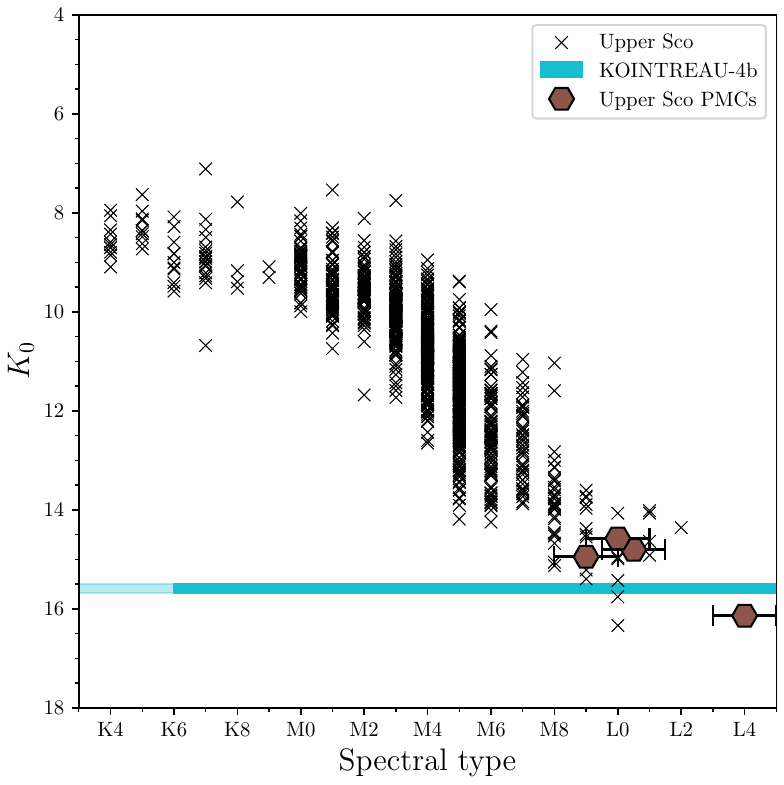}
    \caption{Extinction-corrected apparent $K$-band magnitude as a function of spectral type for our candidate companions, plotted against K4-L3 $\rho$ Oph sources and PMCs (left), and K4-L3 Upper Sco sources and PMCs (right). The $K$-band magnitudes of our two candidate companions are roughly consistent with $>$L0 for KOINTREAU\nobreakdash-\hspace{0pt}3b and L0$\pm$2 for KOINTREAU\nobreakdash-\hspace{0pt}4b. The lighter region of each band indicates the range of spectral types disfavored by the \textit{Gaia} non-detection of each companion, calculated by combining $G-K$ color as a function of spectral type \citep{pecaut2013} with the observed $K$-band magnitude of each companion and \textit{Gaia} $G$-magnitude contrast limits \citep{brandeker2019}. These bands are $\pm1\sigma$ tall. Ophiuchus and Upper Sco PMCs are the same as in Figure~\ref{fig:cmd_byspt}.}
    \label{fig:kmagvsspt}
\end{figure*}

\subsection{Color-magnitude diagram}
\label{sec:cmd}

After dereddening the apparent magnitudes for each object using extinction values derived for the host stars from \citet{esplin2020}, we placed them on a color-magnitude diagram, comparing them to Ophiuchus and Upper Sco sources (Figure~\ref{fig:cmd_byspt}). We also compare the extinction-corrected apparent $K$-band magnitudes of our two companions to the Ophiuchus and Upper Sco populations, plotted as a function of spectral type (Figure~\ref{fig:kmagvsspt}). Given our lack of spectroscopic data for these sources, we use the properties of our two companions compared to the wider Ophiuchus population to approximate their spectral types. We estimate that KOINTREAU\nobreakdash-\hspace{0pt}3b has a spectral type $>$L0, and that KOINTREAU\nobreakdash-\hspace{0pt}4b has a spectral type of L0~$\pm$~2. Comparing KOINTREAU\nobreakdash-\hspace{0pt}4b to the Upper Sco population does not alter our estimate of its spectral type.

\subsection{Companion Masses}
\label{sec:masses}

Once we had verified that these objects are comoving companions to their host stars, we used their absolute magnitudes to estimate masses for KOINTREAU\nobreakdash-\hspace{0pt}3b and KOINTREAU\nobreakdash-\hspace{0pt}4b, adopting the same approach as described in Paper I. In brief, we first computed each object's bolometric luminosity (L$_{\rm bol}$) from the dereddened magnitudes of each companion and the \citet{sanghi2023} bolometric luminosities for young ultracool objects as a function of spectral type (using the estimated spectral types from our photometry for each object). We then linearly interpolated the \texttt{DUSTY} model grid \citep{dusty, dusty2} in log(age), log(L$_{\rm bol}$) and log(mass), and used this to obtain a posterior distribution of log(mass) values given our computed L$_{\rm bol}$ and representative ages. For KOINTREAU\nobreakdash-\hspace{0pt}4b, we did this separately with the L$_{\rm bol}$ values derived from each of its $JHK$ magnitudes, then merged the posteriors from each magnitude. We took the median of the final posterior to estimate the mass of each companion, using the 68\% confidence interval as the uncertainty. 
We note that these mass estimates assume that these two objects experience no significant additional extinction than their host stars (e.g. from an edge-on disk). These masses should be regarded as preliminary until we are able to obtain spectra of these sources to confirm their assumed spectral types and extinctions.

For KOINTREAU\nobreakdash-\hspace{0pt}3b, we adopted the system age cited by \citet{esplin2018} of $2\pm1$ Myr, as this best agrees with other literature estimates of the age of Ophiuchus. We also adopted a nominal spectral type of L$5\pm5$ to allow us to estimate the bolometric luminosity, although we note that the \citet{sanghi2023} bolometric corrections vary little across the L-types (see their Figure 19). At this age and estimated spectral type, our analysis calculates KOINTREAU\nobreakdash-\hspace{0pt}3b to have a mass of $3.4\pm0.7$~M$_{\rm Jup}$. We note that even at the oldest age for this sytem from the literature, our analysis still results in a secure planetary mass for KOINTREAU\nobreakdash-\hspace{0pt}3b ($7.5^{+0.9}_{-1.0}$~M$_{\rm Jup}$ at the \citet{kerr2021} individual stellar age of $11.0^{+2.6}_{-3.1}$ Myr).

Due to the uncertain age of 2M~1626\nobreakdash$-$\hspace{0pt}2625 resulting from its uncertain cluster membership, we repeated the described analysis twice using available literature ages corresponding to membership of both Ophiuchus and Upper Sco. 
This showed that KOINTREAU\nobreakdash-\hspace{0pt}4b could have a mass of either $11.5^{+1.2}_{-1.6}$~M$_{\rm Jup}$ (assuming the \citeauthor{esplin2020} off-cloud Ophiuchus subcluster age of $5\pm1$ Myr) or $15.3^{+0.7}_{-0.8}$~M$_{\rm Jup}$ (assuming the \citeauthor{kerr2021} Upper Sco cluster age of $11.3\pm0.3$ Myr).

The \texttt{DUSTY} models also predict the absolute $K$-band magnitude of sources as a function of their mass and age (i.e. without any reliance on the spectral type of the source, as was the case above when using bolometric corrections). We used this approach to verify the derived masses provided above and obtained consistent results, implying that any error induced by our estimated spectral types is within our derived error bars.

A consensus on the ages of the host stars, as well as spectroscopically-derived spectral types and extinctions for the companions, will be crucial in refining these masses.

\section{Conclusion}

We have discovered new companions to the young stars ISO-Oph~96 and 2MASS J16262785$-$262515. Relative astrometry from combining our multi-year Keck/NIRC2 AO imaging with \textit{Gaia} astrometry for the host stars has established that both companions are bound to their host stars.

We used our photometry to estimate spectral types for both objects by comparison with literature Ophiuchus objects. This gave that KOINTREAU\nobreakdash-\hspace{0pt}3b (the imaged companion to ISO-Oph~96) has a spectral type $>$L0, and that KOINTREAU\nobreakdash-\hspace{0pt}4b (the companion to 2M 1626$-$2625) has a spectral type of L0~$\pm$~2.

We estimate the masses for our new discoveries using the \texttt{DUSTY} evolutionary models and our observed photometry, assuming that the companions experience the same extinction as their host stars. Under this assumption, KOINTREAU\nobreakdash-\hspace{0pt}3b is securely the lowest-mass directly imaged companion in Ophiuchus, with an estimated mass of $3.4\pm0.7$~M$_{\rm Jup}$. The uncertain cluster membership of its host star results in a range of estimated masses for KOINTREAU\nobreakdash-\hspace{0pt}4b which straddle the planetary-mass boundary, from $11.5^{+1.2}_{-1.6}$~M$_{\rm Jup}$ to $15.3^{+0.7}_{-0.8}$~M$_{\rm Jup}$.

Near-infrared spectroscopy of these two companions is necessary to confirm their spectral types and extinctions. This, coupled with more secure ages for their host stars, will enable us to refine their mass estimates.

\begin{acknowledgments}

This research was funded in part by the Gordon and Betty Moore Foundation through grant GBMF8550 to M.~Liu.

This material is based on work supported by the National Science Foundation Graduate Research Fellowship under Grant No.~2139433.

Some of the data presented herein were obtained at Keck Observatory, which is a private 501(c)3 non-profit organization operated as a scientific partnership among the California Institute of Technology, the University of California, and the National Aeronautics and Space Administration. The Observatory was made possible by the generous financial support of the W. M. Keck Foundation.

This work has made use of data from the European Space Agency (ESA) mission
{\it Gaia} (\url{https://www.cosmos.esa.int/gaia}), processed by the {\it Gaia}
Data Processing and Analysis Consortium (DPAC,
\url{https://www.cosmos.esa.int/web/gaia/dpac/consortium}). Funding for the DPAC
has been provided by national institutions, in particular the institutions
participating in the {\it Gaia} Multilateral Agreement.



The authors wish to recognize and acknowledge the very significant cultural role and reverence that the summit of Maunakea has always had within the Native Hawaiian community. We are grateful for the privilege of observing the Universe from a place that is unique in both its astronomical quality and its cultural significance.
\end{acknowledgments}

%

\facility{Keck II (NIRC2)}


\software{astropy \citep{astropy1, astropy2, astropy3}; ccdproc \citep{ccdproc}}



\FloatBarrier

\bibliography{bib}{}

@ARTICLE{service2016,
       author = {{Service}, M. and {Lu}, J.~R. and {Campbell}, R. and {Sitarski}, B.~N. and {Ghez}, A.~M. and {Anderson}, J.},
        title = "{A New Distortion Solution for NIRC2 on the Keck II Telescope}",
      journal = {\pasp},
         year = 2016,
        month = sep,
       volume = {128},
       number = {967},
        pages = {095004},
          doi = {10.1088/1538-3873/128/967/095004},
       adsurl = {https://ui.adsabs.harvard.edu/abs/2016PASP..128i5004S}
}

@misc{ccdproc,
    author       = {Matt Craig and Steve Crawford and Michael Seifert and
                    Thomas Robitaille and Brigitta Sip{\H o}cz and
                    Josh Walawender and Z\`e Vin{\'{\i}}cius and Joe Philip Ninan and Michael Droettboom and Jiyong Youn and
                    Erik Tollerud and Erik Bray and
                    Nathan Walker and VSN Reddy Janga and
                    Connor Stotts and Hans Moritz G{\"u}nther and Evert Rol and
                    Yoonsoo P. Bach and Larry Bradley and Christoph Deil and
                    Adrian Price-Whelan and Kyle Barbary and Anthony Horton and
                    William Schoenell and Nathan Heidt and Forrest Gasdia and
                    Stefan Nelson and Ole Streicher},
    title        = {astropy/ccdproc: v1.3.0.post1},
    month        = dec,
    year         = 2017,
    doi          = {10.5281/zenodo.1069648},
    url          = {https://doi.org/10.5281/zenodo.1069648}
}

@ARTICLE{skrutskie2006,
       author = {{Skrutskie}, M.~F. and {Cutri}, R.~M. and {Stiening}, R. and {Weinberg}, M.~D. and {Schneider}, S. and {Carpenter}, J.~M. and {Beichman}, C. and {Capps}, R. and {Chester}, T. and {Elias}, J. and {Huchra}, J. and {Liebert}, J. and {Lonsdale}, C. and {Monet}, D.~G. and {Price}, S. and {Seitzer}, P. and {Jarrett}, T. and {Kirkpatrick}, J.~D. and {Gizis}, J.~E. and {Howard}, E. and {Evans}, T. and {Fowler}, J. and {Fullmer}, L. and {Hurt}, R. and {Light}, R. and {Kopan}, E.~L. and {Marsh}, K.~A. and {McCallon}, H.~L. and {Tam}, R. and {Van Dyk}, S. and {Wheelock}, S.},
        title = "{The Two Micron All Sky Survey (2MASS)}",
      journal = {\aj},
         year = 2006,
        month = feb,
       volume = {131},
       number = {2},
        pages = {1163-1183},
          doi = {10.1086/498708},
       adsurl = {https://ui.adsabs.harvard.edu/abs/2006AJ....131.1163S}
}

@ARTICLE{sanghi2023,
       author = {{Sanghi}, Aniket and {Liu}, Michael C. and {Best}, William M.~J. and {Dupuy}, Trent J. and {Siverd}, Robert J. and {Zhang}, Zhoujian and {Hurt}, Spencer A. and {Magnier}, Eugene A. and {Aller}, Kimberly M. and {Deacon}, Niall R.},
        title = "{The Hawaii Infrared Parallax Program. VI. The Fundamental Properties of 1000+ Ultracool Dwarfs and Planetary-mass Objects Using Optical to Mid-infrared Spectral Energy Distributions and Comparison to BT-Settl and ATMO 2020 Model Atmospheres}",
      journal = {\apj},
         year = 2023,
        month = dec,
       volume = {959},
       number = {1},
          eid = {63},
        pages = {63},
          doi = {10.3847/1538-4357/acff66},
archivePrefix = {arXiv},
       eprint = {2309.03082},
 primaryClass = {astro-ph.SR},
       adsurl = {https://ui.adsabs.harvard.edu/abs/2023ApJ...959...63S}
}

@ARTICLE{jonsson2020,
       author = {{J{\"o}nsson}, Henrik and {Holtzman}, Jon A. and {Allende Prieto}, Carlos and {Cunha}, Katia and {Garc{\'\i}a-Hern{\'a}ndez}, D.~A. and {Hasselquist}, Sten and {Masseron}, Thomas and {Osorio}, Yeisson and {Shetrone}, Matthew and {Smith}, Verne and {Stringfellow}, Guy S. and {Bizyaev}, Dmitry and {Edvardsson}, Bengt and {Majewski}, Steven R. and {M{\'e}sz{\'a}ros}, Szabolcs and {Souto}, Diogo and {Zamora}, Olga and {Beaton}, Rachael L. and {Bovy}, Jo and {Donor}, John and {Pinsonneault}, Marc H. and {Poovelil}, Vijith Jacob and {Sobeck}, Jennifer},
        title = "{APOGEE Data and Spectral Analysis from SDSS Data Release 16: Seven Years of Observations Including First Results from APOGEE-South}",
      journal = {\aj},
         year = 2020,
        month = sep,
       volume = {160},
       number = {3},
          eid = {120},
        pages = {120},
          doi = {10.3847/1538-3881/aba592},
archivePrefix = {arXiv},
       eprint = {2007.05537},
 primaryClass = {astro-ph.GA},
       adsurl = {https://ui.adsabs.harvard.edu/abs/2020AJ....160..120J}
}

@ARTICLE{cutri2003,
       author = {{Cutri}, R.~M. and {Skrutskie}, M.~F. and {van Dyk}, S. and {Beichman}, C.~A. and {Carpenter}, J.~M. and {Chester}, T. and {Cambresy}, L. and {Evans}, T. and {Fowler}, J. and {Gizis}, J. and {Howard}, E. and {Huchra}, J. and {Jarrett}, T. and {Kopan}, E.~L. and {Kirkpatrick}, J.~D. and {Light}, R.~M. and {Marsh}, K.~A. and {McCallon}, H. and {Schneider}, S. and {Stiening}, R. and {Sykes}, M. and {Weinberg}, M. and {Wheaton}, W.~A. and {Wheelock}, S. and {Zacarias}, N.},
        title = "{VizieR Online Data Catalog: 2MASS All-Sky Catalog of Point Sources (Cutri+ 2003)}",
      journal = {VizieR Online Data Catalog},
         year = 2003,
        month = jun,
          eid = {II/246},
        pages = {II/246},
       adsurl = {https://ui.adsabs.harvard.edu/abs/2003yCat.2246....0C}
}

@ARTICLE{gaiadr3,
       author = {{Gaia Collaboration} and {Vallenari}, A. and {Brown}, A.~G.~A. and {Prusti}, T. and {de Bruijne}, J.~H.~J. and {Arenou}, F. and {Babusiaux}, C. and {Biermann}, M. and {Creevey}, O.~L. and {Ducourant}, C. and {Evans}, D.~W. and {Eyer}, L. and {Guerra}, R. and {Hutton}, A. and {Jordi}, C. and {Klioner}, S.~A. and {Lammers}, U.~L. and {Lindegren}, L. and {Luri}, X. and {Mignard}, F. and {Panem}, C. and {Pourbaix}, D. and {Randich}, S. and {Sartoretti}, P. and {Soubiran}, C. and {Tanga}, P. and {Walton}, N.~A. and {Bailer-Jones}, C.~A.~L. and {Bastian}, U. and {Drimmel}, R. and {Jansen}, F. and {Katz}, D. and {Lattanzi}, M.~G. and {van Leeuwen}, F. and {Bakker}, J. and {Cacciari}, C. and {Casta{\~n}eda}, J. and {De Angeli}, F. and {Fabricius}, C. and {Fouesneau}, M. and {Fr{\'e}mat}, Y. and {Galluccio}, L. and {Guerrier}, A. and {Heiter}, U. and {Masana}, E. and {Messineo}, R. and {Mowlavi}, N. and {Nicolas}, C. and {Nienartowicz}, K. and {Pailler}, F. and {Panuzzo}, P. and {Riclet}, F. and {Roux}, W. and {Seabroke}, G.~M. and {Sordo}, R. and {Th{\'e}venin}, F. and {Gracia-Abril}, G. and {Portell}, J. and {Teyssier}, D. and {Altmann}, M. and {Andrae}, R. and {Audard}, M. and {Bellas-Velidis}, I. and {Benson}, K. and {Berthier}, J. and {Blomme}, R. and {Burgess}, P.~W. and {Busonero}, D. and {Busso}, G. and {C{\'a}novas}, H. and {Carry}, B. and {Cellino}, A. and {Cheek}, N. and {Clementini}, G. and {Damerdji}, Y. and {Davidson}, M. and {de Teodoro}, P. and {Nu{\~n}ez Campos}, M. and {Delchambre}, L. and {Dell'Oro}, A. and {Esquej}, P. and {Fern{\'a}ndez-Hern{\'a}ndez}, J. and {Fraile}, E. and {Garabato}, D. and {Garc{\'\i}a-Lario}, P. and {Gosset}, E. and {Haigron}, R. and {Halbwachs}, J. -L. and {Hambly}, N.~C. and {Harrison}, D.~L. and {Hern{\'a}ndez}, J. and {Hestroffer}, D. and {Hodgkin}, S.~T. and {Holl}, B. and {Jan{\ss}en}, K. and {Jevardat de Fombelle}, G. and {Jordan}, S. and {Krone-Martins}, A. and {Lanzafame}, A.~C. and {L{\"o}ffler}, W. and {Marchal}, O. and {Marrese}, P.~M. and {Moitinho}, A. and {Muinonen}, K. and {Osborne}, P. and {Pancino}, E. and {Pauwels}, T. and {Recio-Blanco}, A. and {Reyl{\'e}}, C. and {Riello}, M. and {Rimoldini}, L. and {Roegiers}, T. and {Rybizki}, J. and {Sarro}, L.~M. and {Siopis}, C. and {Smith}, M. and {Sozzetti}, A. and {Utrilla}, E. and {van Leeuwen}, M. and {Abbas}, U. and {{\'A}brah{\'a}m}, P. and {Abreu Aramburu}, A. and {Aerts}, C. and {Aguado}, J.~J. and {Ajaj}, M. and {Aldea-Montero}, F. and {Altavilla}, G. and {{\'A}lvarez}, M.~A. and {Alves}, J. and {Anders}, F. and {Anderson}, R.~I. and {Anglada Varela}, E. and {Antoja}, T. and {Baines}, D. and {Baker}, S.~G. and {Balaguer-N{\'u}{\~n}ez}, L. and {Balbinot}, E. and {Balog}, Z. and {Barache}, C. and {Barbato}, D. and {Barros}, M. and {Barstow}, M.~A. and {Bartolom{\'e}}, S. and {Bassilana}, J. -L. and {Bauchet}, N. and {Becciani}, U. and {Bellazzini}, M. and {Berihuete}, A. and {Bernet}, M. and {Bertone}, S. and {Bianchi}, L. and {Binnenfeld}, A. and {Blanco-Cuaresma}, S. and {Blazere}, A. and {Boch}, T. and {Bombrun}, A. and {Bossini}, D. and {Bouquillon}, S. and {Bragaglia}, A. and {Bramante}, L. and {Breedt}, E. and {Bressan}, A. and {Brouillet}, N. and {Brugaletta}, E. and {Bucciarelli}, B. and {Burlacu}, A. and {Butkevich}, A.~G. and {Buzzi}, R. and {Caffau}, E. and {Cancelliere}, R. and {Cantat-Gaudin}, T. and {Carballo}, R. and {Carlucci}, T. and {Carnerero}, M.~I. and {Carrasco}, J.~M. and {Casamiquela}, L. and {Castellani}, M. and {Castro-Ginard}, A. and {Chaoul}, L. and {Charlot}, P. and {Chemin}, L. and {Chiaramida}, V. and {Chiavassa}, A. and {Chornay}, N. and {Comoretto}, G. and {Contursi}, G. and {Cooper}, W.~J. and {Cornez}, T. and {Cowell}, S. and {Crifo}, F. and {Cropper}, M. and {Crosta}, M. and {Crowley}, C. and {Dafonte}, C. and {Dapergolas}, A. and {David}, M. and {David}, P. and {de Laverny}, P. and {De Luise}, F. and {De March}, R. and {De Ridder}, J. and {de Souza}, R. and {de Torres}, A. and {del Peloso}, E.~F. and {del Pozo}, E. and {Delbo}, M. and {Delgado}, A. and {Delisle}, J. -B. and {Demouchy}, C. and {Dharmawardena}, T.~E. and {Di Matteo}, P. and {Diakite}, S. and {Diener}, C. and {Distefano}, E. and {Dolding}, C. and {Edvardsson}, B. and {Enke}, H. and {Fabre}, C. and {Fabrizio}, M. and {Faigler}, S. and {Fedorets}, G. and {Fernique}, P. and {Fienga}, A. and {Figueras}, F. and {Fournier}, Y. and {Fouron}, C. and {Fragkoudi}, F. and {Gai}, M. and {Garcia-Gutierrez}, A. and {Garcia-Reinaldos}, M. and {Garc{\'\i}a-Torres}, M. and {Garofalo}, A. and {Gavel}, A. and {Gavras}, P. and {Gerlach}, E. and {Geyer}, R. and {Giacobbe}, P. and {Gilmore}, G. and {Girona}, S. and {Giuffrida}, G. and {Gomel}, R. and {Gomez}, A. and {Gonz{\'a}lez-N{\'u}{\~n}ez}, J. and {Gonz{\'a}lez-Santamar{\'\i}a}, I. and {Gonz{\'a}lez-Vidal}, J.~J. and {Granvik}, M. and {Guillout}, P. and {Guiraud}, J. and {Guti{\'e}rrez-S{\'a}nchez}, R. and {Guy}, L.~P. and {Hatzidimitriou}, D. and {Hauser}, M. and {Haywood}, M. and {Helmer}, A. and {Helmi}, A. and {Sarmiento}, M.~H. and {Hidalgo}, S.~L. and {Hilger}, T. and {H{\l}adczuk}, N. and {Hobbs}, D. and {Holland}, G. and {Huckle}, H.~E. and {Jardine}, K. and {Jasniewicz}, G. and {Jean-Antoine Piccolo}, A. and {Jim{\'e}nez-Arranz}, {\'O}. and {Jorissen}, A. and {Juaristi Campillo}, J. and {Julbe}, F. and {Karbevska}, L. and {Kervella}, P. and {Khanna}, S. and {Kontizas}, M. and {Kordopatis}, G. and {Korn}, A.~J. and {K{\'o}sp{\'a}l}, {\'A}. and {Kostrzewa-Rutkowska}, Z. and {Kruszy{\'n}ska}, K. and {Kun}, M. and {Laizeau}, P. and {Lambert}, S. and {Lanza}, A.~F. and {Lasne}, Y. and {Le Campion}, J. -F. and {Lebreton}, Y. and {Lebzelter}, T. and {Leccia}, S. and {Leclerc}, N. and {Lecoeur-Taibi}, I. and {Liao}, S. and {Licata}, E.~L. and {Lindstr{\o}m}, H.~E.~P. and {Lister}, T.~A. and {Livanou}, E. and {Lobel}, A. and {Lorca}, A. and {Loup}, C. and {Madrero Pardo}, P. and {Magdaleno Romeo}, A. and {Managau}, S. and {Mann}, R.~G. and {Manteiga}, M. and {Marchant}, J.~M. and {Marconi}, M. and {Marcos}, J. and {Marcos Santos}, M.~M.~S. and {Mar{\'\i}n Pina}, D. and {Marinoni}, S. and {Marocco}, F. and {Marshall}, D.~J. and {Martin Polo}, L. and {Mart{\'\i}n-Fleitas}, J.~M. and {Marton}, G. and {Mary}, N. and {Masip}, A. and {Massari}, D. and {Mastrobuono-Battisti}, A. and {Mazeh}, T. and {McMillan}, P.~J. and {Messina}, S. and {Michalik}, D. and {Millar}, N.~R. and {Mints}, A. and {Molina}, D. and {Molinaro}, R. and {Moln{\'a}r}, L. and {Monari}, G. and {Mongui{\'o}}, M. and {Montegriffo}, P. and {Montero}, A. and {Mor}, R. and {Mora}, A. and {Morbidelli}, R. and {Morel}, T. and {Morris}, D. and {Muraveva}, T. and {Murphy}, C.~P. and {Musella}, I. and {Nagy}, Z. and {Noval}, L. and {Oca{\~n}a}, F. and {Ogden}, A. and {Ordenovic}, C. and {Osinde}, J.~O. and {Pagani}, C. and {Pagano}, I. and {Palaversa}, L. and {Palicio}, P.~A. and {Pallas-Quintela}, L. and {Panahi}, A. and {Payne-Wardenaar}, S. and {Pe{\~n}alosa Esteller}, X. and {Penttil{\"a}}, A. and {Pichon}, B. and {Piersimoni}, A.~M. and {Pineau}, F. -X. and {Plachy}, E. and {Plum}, G. and {Poggio}, E. and {Pr{\v{s}}a}, A. and {Pulone}, L. and {Racero}, E. and {Ragaini}, S. and {Rainer}, M. and {Raiteri}, C.~M. and {Rambaux}, N. and {Ramos}, P. and {Ramos-Lerate}, M. and {Re Fiorentin}, P. and {Regibo}, S. and {Richards}, P.~J. and {Rios Diaz}, C. and {Ripepi}, V. and {Riva}, A. and {Rix}, H. -W. and {Rixon}, G. and {Robichon}, N. and {Robin}, A.~C. and {Robin}, C. and {Roelens}, M. and {Rogues}, H.~R.~O. and {Rohrbasser}, L. and {Romero-G{\'o}mez}, M. and {Rowell}, N. and {Royer}, F. and {Ruz Mieres}, D. and {Rybicki}, K.~A. and {Sadowski}, G. and {S{\'a}ez N{\'u}{\~n}ez}, A. and {Sagrist{\`a} Sell{\'e}s}, A. and {Sahlmann}, J. and {Salguero}, E. and {Samaras}, N. and {Sanchez Gimenez}, V. and {Sanna}, N. and {Santove{\~n}a}, R. and {Sarasso}, M. and {Schultheis}, M. and {Sciacca}, E. and {Segol}, M. and {Segovia}, J.~C. and {S{\'e}gransan}, D. and {Semeux}, D. and {Shahaf}, S. and {Siddiqui}, H.~I. and {Siebert}, A. and {Siltala}, L. and {Silvelo}, A. and {Slezak}, E. and {Slezak}, I. and {Smart}, R.~L. and {Snaith}, O.~N. and {Solano}, E. and {Solitro}, F. and {Souami}, D. and {Souchay}, J. and {Spagna}, A. and {Spina}, L. and {Spoto}, F. and {Steele}, I.~A. and {Steidelm{\"u}ller}, H. and {Stephenson}, C.~A. and {S{\"u}veges}, M. and {Surdej}, J. and {Szabados}, L. and {Szegedi-Elek}, E. and {Taris}, F. and {Taylor}, M.~B. and {Teixeira}, R. and {Tolomei}, L. and {Tonello}, N. and {Torra}, F. and {Torra}, J. and {Torralba Elipe}, G. and {Trabucchi}, M. and {Tsounis}, A.~T. and {Turon}, C. and {Ulla}, A. and {Unger}, N. and {Vaillant}, M.~V. and {van Dillen}, E. and {van Reeven}, W. and {Vanel}, O. and {Vecchiato}, A. and {Viala}, Y. and {Vicente}, D. and {Voutsinas}, S. and {Weiler}, M. and {Wevers}, T. and {Wyrzykowski}, {\L}. and {Yoldas}, A. and {Yvard}, P. and {Zhao}, H. and {Zorec}, J. and {Zucker}, S. and {Zwitter}, T.},
        title = "{Gaia Data Release 3. Summary of the content and survey properties}",
      journal = {\aap},
         year = 2023,
        month = jun,
       volume = {674},
          eid = {A1},
        pages = {A1},
          doi = {10.1051/0004-6361/202243940},
archivePrefix = {arXiv},
       eprint = {2208.00211},
 primaryClass = {astro-ph.GA},
       adsurl = {https://ui.adsabs.harvard.edu/abs/2023A&A...674A...1G}
}

@ARTICLE{dusty,
       author = {{Chabrier}, G. and {Baraffe}, I. and {Allard}, F. and {Hauschildt}, P.},
        title = "{Evolutionary Models for Very Low-Mass Stars and Brown Dwarfs with Dusty Atmospheres}",
      journal = {\apj},
         year = 2000,
        month = oct,
       volume = {542},
       number = {1},
        pages = {464-472},
          doi = {10.1086/309513},
archivePrefix = {arXiv},
       eprint = {astro-ph/0005557},
 primaryClass = {astro-ph},
       adsurl = {https://ui.adsabs.harvard.edu/abs/2000ApJ...542..464C}
}

@ARTICLE{dusty2,
       author = {{Baraffe}, I. and {Chabrier}, G. and {Allard}, F. and {Hauschildt}, P.~H.},
        title = "{Evolutionary models for low-mass stars and brown dwarfs: Uncertainties and limits at very young ages}",
      journal = {\aap},
         year = 2002,
        month = feb,
       volume = {382},
        pages = {563-572},
          doi = {10.1051/0004-6361:20011638},
archivePrefix = {arXiv},
       eprint = {astro-ph/0111385},
 primaryClass = {astro-ph},
       adsurl = {https://ui.adsabs.harvard.edu/abs/2002A&A...382..563B}
}

@ARTICLE{kerr2021,
       author = {{Kerr}, Ronan M.~P. and {Rizzuto}, Aaron C. and {Kraus}, Adam L. and {Offner}, Stella S.~R.},
        title = "{Stars with Photometrically Young Gaia Luminosities Around the Solar System (SPYGLASS). I. Mapping Young Stellar Structures and Their Star Formation Histories}",
      journal = {\apj},
         year = 2021,
        month = aug,
       volume = {917},
       number = {1},
          eid = {23},
        pages = {23},
          doi = {10.3847/1538-4357/ac0251},
archivePrefix = {arXiv},
       eprint = {2105.09338},
 primaryClass = {astro-ph.GA},
       adsurl = {https://ui.adsabs.harvard.edu/abs/2021ApJ...917...23K}
}

@ARTICLE{pywfsjatis,
       author = {{Bond}, Charlotte Z. and {Cetre}, Sylvain and {Lilley}, Scott and {Wizinowich}, Peter and {Mawet}, Dimitri and {Chun}, Mark and {Wetherell}, Edward and {Jacobson}, Shane and {Lockhart}, Charles and {Warmbier}, Eric and {Ragland}, Sam and {Alvarez}, Carlos and {Guyon}, Olivier and {Goebel}, Sean and {Delorme}, Jacques-Robert and {Jovanovic}, Nemanja and {Hall}, Donald N. and {Wallace}, James K. and {Taheri}, Mojtaba and {Plantet}, Cedric and {Chambouleyron}, Vincent},
        title = "{Adaptive optics with an infrared pyramid wavefront sensor at Keck}",
      journal = {Journal of Astronomical Telescopes, Instruments, and Systems},
         year = 2020,
        month = jul,
       volume = {6},
          eid = {039003},
        pages = {039003},
          doi = {10.1117/1.JATIS.6.3.039003},
       adsurl = {https://ui.adsabs.harvard.edu/abs/2020JATIS...6c9003B}
}

@ARTICLE{bhac2015,
       author = {{Baraffe}, Isabelle and {Homeier}, Derek and {Allard}, France and {Chabrier}, Gilles},
        title = "{New evolutionary models for pre-main sequence and main sequence low-mass stars down to the hydrogen-burning limit}",
      journal = {\aap},
         year = 2015,
        month = may,
       volume = {577},
          eid = {A42},
        pages = {A42},
          doi = {10.1051/0004-6361/201425481},
archivePrefix = {arXiv},
       eprint = {1503.04107},
 primaryClass = {astro-ph.SR},
       adsurl = {https://ui.adsabs.harvard.edu/abs/2015A&A...577A..42B}
}

@ARTICLE{wizinowich2000,
       author = {{Wizinowich}, P. and {Acton}, D.~S. and {Shelton}, C. and {Stomski}, P. and {Gathright}, J. and {Ho}, K. and {Lupton}, W. and {Tsubota}, K. and {Lai}, O. and {Max}, C. and {Brase}, J. and {An}, J. and {Avicola}, K. and {Olivier}, S. and {Gavel}, D. and {Macintosh}, B. and {Ghez}, A. and {Larkin}, J.},
        title = "{First Light Adaptive Optics Images from the Keck II Telescope: A New Era of High Angular Resolution Imagery}",
      journal = {\pasp},
         year = 2000,
        month = mar,
       volume = {112},
       number = {769},
        pages = {315-319},
          doi = {10.1086/316543},
       adsurl = {https://ui.adsabs.harvard.edu/abs/2000PASP..112..315W}
}

@ARTICLE{bpmgage,
       author = {{Lee}, Rena A. and {Gaidos}, Eric and {van Saders}, Jennifer and {Feiden}, Gregory A. and {Gagn{\'e}}, Jonathan},
        title = "{Revisiting the membership, multiplicity, and age of the Beta Pictoris Moving Group in the Gaia era}",
      journal = {\mnras},
         year = 2024,
        month = mar,
       volume = {528},
       number = {3},
        pages = {4760-4774},
          doi = {10.1093/mnras/stae007},
archivePrefix = {arXiv},
       eprint = {2312.15792},
 primaryClass = {astro-ph.SR},
       adsurl = {https://ui.adsabs.harvard.edu/abs/2024MNRAS.528.4760L}
}

@ARTICLE{esplin2020,
       author = {{Esplin}, T.~L. and {Luhman}, K.~L.},
        title = "{A Survey for New Stars and Brown Dwarfs in the Ophiuchus Star-forming Complex}",
      journal = {\aj},
         year = 2020,
        month = jun,
       volume = {159},
       number = {6},
          eid = {282},
        pages = {282},
          doi = {10.3847/1538-3881/ab8dbd},
archivePrefix = {arXiv},
       eprint = {2005.10096},
 primaryClass = {astro-ph.SR},
       adsurl = {https://ui.adsabs.harvard.edu/abs/2020AJ....159..282E}
}

@ARTICLE{padgett2008,
       author = {{Padgett}, Deborah L. and {Rebull}, Luisa M. and {Stapelfeldt}, Karl R. and {Chapman}, Nicholas L. and {Lai}, Shih-Ping and {Mundy}, Lee G. and {Evans}, II, Neal J. and {Brooke}, Timothy Y. and {Cieza}, Lucas A. and {Spiesman}, William J. and {Noriega-Crespo}, Alberto and {McCabe}, Caer-Eve and {Allen}, Lori E. and {Blake}, Geoffrey A. and {Harvey}, Paul M. and {Huard}, Tracy L. and {J{\o}rgensen}, Jes K. and {Koerner}, David W. and {Myers}, Philip C. and {Sargent}, Annelia I. and {Teuben}, Peter and {van Dishoeck}, Ewine F. and {Wahhaj}, Zahed and {Young}, Kaisa E.},
        title = "{The Spitzer c2d Survey of Large, Nearby, Interstellar Clouds. VII. Ophiuchus Observed with MIPS}",
      journal = {\apj},
         year = 2008,
        month = jan,
       volume = {672},
       number = {2},
        pages = {1013-1037},
          doi = {10.1086/523883},
archivePrefix = {arXiv},
       eprint = {0709.3492},
 primaryClass = {astro-ph},
       adsurl = {https://ui.adsabs.harvard.edu/abs/2008ApJ...672.1013P}
}

@ARTICLE{luhman2020,
       author = {{Luhman}, K.~L. and {Esplin}, T.~L.},
        title = "{Refining the Census of the Upper Scorpius Association with Gaia}",
      journal = {\aj},
         year = 2020,
        month = jul,
       volume = {160},
       number = {1},
          eid = {44},
        pages = {44},
          doi = {10.3847/1538-3881/ab9599},
archivePrefix = {arXiv},
       eprint = {2005.10128},
 primaryClass = {astro-ph.SR},
       adsurl = {https://ui.adsabs.harvard.edu/abs/2020AJ....160...44L}
}

@ARTICLE{brandeker2019,
       author = {{Brandeker}, Alexis and {Cataldi}, Gianni},
        title = "{Contrast sensitivities in the Gaia Data Release 2}",
      journal = {\aap},
         year = 2019,
        month = jan,
       volume = {621},
          eid = {A86},
        pages = {A86},
          doi = {10.1051/0004-6361/201834321},
archivePrefix = {arXiv},
       eprint = {1811.05301},
 primaryClass = {astro-ph.IM},
       adsurl = {https://ui.adsabs.harvard.edu/abs/2019A&A...621A..86B}
}

@ARTICLE{besancon,
       author = {{Robin}, A.~C. and {Reyl{\'e}}, C. and {Derri{\`e}re}, S. and {Picaud}, S.},
        title = "{A synthetic view on structure and evolution of the Milky Way}",
      journal = {\aap},
         year = 2003,
        month = oct,
       volume = {409},
        pages = {523-540},
          doi = {10.1051/0004-6361:20031117},
       adsurl = {https://ui.adsabs.harvard.edu/abs/2003A&A...409..523R}
}

@ARTICLE{natta2006,
       author = {{Natta}, A. and {Testi}, L. and {Randich}, S.},
        title = "{Accretion in the {\ensuremath{\rho}}-Ophiuchi pre-main sequence stars}",
      journal = {\aap},
         year = 2006,
        month = jun,
       volume = {452},
       number = {1},
        pages = {245-252},
          doi = {10.1051/0004-6361:20054706},
archivePrefix = {arXiv},
       eprint = {astro-ph/0602618},
 primaryClass = {astro-ph},
       adsurl = {https://ui.adsabs.harvard.edu/abs/2006A&A...452..245N}
}

@ARTICLE{pecaut2013,
       author = {{Pecaut}, Mark J. and {Mamajek}, Eric E.},
        title = "{Intrinsic Colors, Temperatures, and Bolometric Corrections of Pre-main-sequence Stars}",
      journal = {\apjs},
         year = 2013,
        month = sep,
       volume = {208},
       number = {1},
          eid = {9},
        pages = {9},
          doi = {10.1088/0067-0049/208/1/9},
archivePrefix = {arXiv},
       eprint = {1307.2657},
 primaryClass = {astro-ph.SR},
       adsurl = {https://ui.adsabs.harvard.edu/abs/2013ApJS..208....9P}
}

@ARTICLE{tokunaga2002,
       author = {{Tokunaga}, A.~T. and {Simons}, D.~A. and {Vacca}, W.~D.},
        title = "{The Mauna Kea Observatories Near-Infrared Filter Set. II. Specifications for a New JHKL'M' Filter Set for Infrared Astronomy}",
      journal = {\pasp},
         year = 2002,
        month = feb,
       volume = {114},
       number = {792},
        pages = {180-186},
          doi = {10.1086/338545},
archivePrefix = {arXiv},
       eprint = {astro-ph/0110593},
 primaryClass = {astro-ph},
       adsurl = {https://ui.adsabs.harvard.edu/abs/2002PASP..114..180T}
}

@ARTICLE{simons2002,
       author = {{Simons}, D.~A. and {Tokunaga}, A.},
        title = "{The Mauna Kea Observatories Near-Infrared Filter Set. I. Defining Optimal 1-5 Micron Bandpasses}",
      journal = {\pasp},
         year = 2002,
        month = feb,
       volume = {114},
       number = {792},
        pages = {169-179},
          doi = {10.1086/338544},
archivePrefix = {arXiv},
       eprint = {astro-ph/0110594},
 primaryClass = {astro-ph},
       adsurl = {https://ui.adsabs.harvard.edu/abs/2002PASP..114..169S}
}

@ARTICLE{wizinowich2006,
       author = {{Wizinowich}, Peter L. and {Le Mignant}, David and {Bouchez}, Antonin H. and {Campbell}, Randy D. and {Chin}, Jason C.~Y. and {Contos}, Adam R. and {van Dam}, Marcos A. and {Hartman}, Scott K. and {Johansson}, Erik M. and {Lafon}, Robert E. and {Lewis}, Hilton and {Stomski}, Paul J. and {Summers}, Douglas M. and {Brown}, Curtis G. and {Danforth}, Pamela M. and {Max}, Claire E. and {Pennington}, Deanna M.},
        title = "{The W. M. Keck Observatory Laser Guide Star Adaptive Optics System: Overview}",
      journal = {\pasp},
         year = 2006,
        month = feb,
       volume = {118},
       number = {840},
        pages = {297-309},
          doi = {10.1086/499290},
       adsurl = {https://ui.adsabs.harvard.edu/abs/2006PASP..118..297W}
}

@ARTICLE{vandam2006,
       author = {{van Dam}, Marcos A. and {Bouchez}, Antonin H. and {Le Mignant}, David and {Johansson}, Erik M. and {Wizinowich}, Peter L. and {Campbell}, Randy D. and {Chin}, Jason C.~Y. and {Hartman}, Scott K. and {Lafon}, Robert E. and {Stomski}, Jr., Paul J. and {Summers}, Douglas M.},
        title = "{The W. M. Keck Observatory Laser Guide Star Adaptive Optics System: Performance Characterization}",
      journal = {\pasp},
         year = 2006,
        month = feb,
       volume = {118},
       number = {840},
        pages = {310-318},
          doi = {10.1086/499498},
       adsurl = {https://ui.adsabs.harvard.edu/abs/2006PASP..118..310V}
}

@ARTICLE{irac,
       author = {{Fazio}, G.~G. and {Hora}, J.~L. and {Allen}, L.~E. and {Ashby}, M.~L.~N. and {Barmby}, P. and {Deutsch}, L.~K. and {Huang}, J. -S. and {Kleiner}, S. and {Marengo}, M. and {Megeath}, S.~T. and {Melnick}, G.~J. and {Pahre}, M.~A. and {Patten}, B.~M. and {Polizotti}, J. and {Smith}, H.~A. and {Taylor}, R.~S. and {Wang}, Z. and {Willner}, S.~P. and {Hoffmann}, W.~F. and {Pipher}, J.~L. and {Forrest}, W.~J. and {McMurty}, C.~W. and {McCreight}, C.~R. and {McKelvey}, M.~E. and {McMurray}, R.~E. and {Koch}, D.~G. and {Moseley}, S.~H. and {Arendt}, R.~G. and {Mentzell}, J.~E. and {Marx}, C.~T. and {Losch}, P. and {Mayman}, P. and {Eichhorn}, W. and {Krebs}, D. and {Jhabvala}, M. and {Gezari}, D.~Y. and {Fixsen}, D.~J. and {Flores}, J. and {Shakoorzadeh}, K. and {Jungo}, R. and {Hakun}, C. and {Workman}, L. and {Karpati}, G. and {Kichak}, R. and {Whitley}, R. and {Mann}, S. and {Tollestrup}, E.~V. and {Eisenhardt}, P. and {Stern}, D. and {Gorjian}, V. and {Bhattacharya}, B. and {Carey}, S. and {Nelson}, B.~O. and {Glaccum}, W.~J. and {Lacy}, M. and {Lowrance}, P.~J. and {Laine}, S. and {Reach}, W.~T. and {Stauffer}, J.~A. and {Surace}, J.~A. and {Wilson}, G. and {Wright}, E.~L. and {Hoffman}, A. and {Domingo}, G. and {Cohen}, M.},
        title = "{The Infrared Array Camera (IRAC) for the Spitzer Space Telescope}",
      journal = {\apjs},
         year = 2004,
        month = sep,
       volume = {154},
       number = {1},
        pages = {10-17},
          doi = {10.1086/422843},
archivePrefix = {arXiv},
       eprint = {astro-ph/0405616},
 primaryClass = {astro-ph},
       adsurl = {https://ui.adsabs.harvard.edu/abs/2004ApJS..154...10F}
}

@ARTICLE{zhang2021,
       author = {{Zhang}, Zhoujian and {Liu}, Michael C. and {Claytor}, Zachary R. and {Best}, William M.~J. and {Dupuy}, Trent J. and {Siverd}, Robert J.},
        title = "{The Second Discovery from the COCONUTS Program: A Cold Wide-orbit Exoplanet around a Young Field M Dwarf at 10.9 pc}",
      journal = {\apjl},
         year = 2021,
        month = aug,
       volume = {916},
       number = {2},
          eid = {L11},
        pages = {L11},
          doi = {10.3847/2041-8213/ac1123},
archivePrefix = {arXiv},
       eprint = {2107.02805},
 primaryClass = {astro-ph.EP},
       adsurl = {https://ui.adsabs.harvard.edu/abs/2021ApJ...916L..11Z}
}

@ARTICLE{lagrange2009,
       author = {{Lagrange}, A.-M. and {Gratadour}, D. and {Chauvin}, G. and {Fusco}, T. and {Ehrenreich}, D. and {Mouillet}, D. and {Rousset}, G. and {Rouan}, D. and {Allard}, F. and {Gendron}, {\'E}. and {Charton}, J. and {Mugnier}, L. and {Rabou}, P. and {Montri}, J. and {Lacombe}, F.},
        title = "{A probable giant planet imaged in the {\ensuremath{\beta}} Pictoris disk. VLT/NaCo deep L'-band imaging}",
      journal = {\aap},
         year = 2009,
        month = jan,
       volume = {493},
       number = {2},
        pages = {L21-L25},
          doi = {10.1051/0004-6361:200811325},
archivePrefix = {arXiv},
       eprint = {0811.3583},
 primaryClass = {astro-ph},
       adsurl = {https://ui.adsabs.harvard.edu/abs/2009A&A...493L..21L}
}

@ARTICLE{naud2014,
       author = {{Naud}, Marie-Eve and {Artigau}, {\'E}tienne and {Malo}, Lison and {Albert}, Lo{\"\i}c and {Doyon}, Ren{\'e} and {Lafreni{\`e}re}, David and {Gagn{\'e}}, Jonathan and {Saumon}, Didier and {Morley}, Caroline V. and {Allard}, France and {Homeier}, Derek and {Beichman}, Charles A. and {Gelino}, Christopher R. and {Boucher}, Anne},
        title = "{Discovery of a Wide Planetary-mass Companion to the Young M3 Star GU Psc}",
      journal = {\apj},
         year = 2014,
        month = may,
       volume = {787},
       number = {1},
          eid = {5},
        pages = {5},
          doi = {10.1088/0004-637X/787/1/5},
archivePrefix = {arXiv},
       eprint = {1405.2932},
 primaryClass = {astro-ph.EP},
       adsurl = {https://ui.adsabs.harvard.edu/abs/2014ApJ...787....5N}
}

@ARTICLE{bowler2016,
       author = {{Bowler}, Brendan P.},
        title = "{Imaging Extrasolar Giant Planets}",
      journal = {\pasp},
         year = 2016,
        month = oct,
       volume = {128},
       number = {968},
        pages = {102001},
          doi = {10.1088/1538-3873/128/968/102001},
archivePrefix = {arXiv},
       eprint = {1605.02731},
 primaryClass = {astro-ph.EP},
       adsurl = {https://ui.adsabs.harvard.edu/abs/2016PASP..128j2001B}
}

@ARTICLE{marois2008,
       author = {{Marois}, Christian and {Macintosh}, Bruce and {Barman}, Travis and {Zuckerman}, B. and {Song}, Inseok and {Patience}, Jennifer and {Lafreni{\`e}re}, David and {Doyon}, Ren{\'e}},
        title = "{Direct Imaging of Multiple Planets Orbiting the Star HR 8799}",
      journal = {Science},
         year = 2008,
        month = nov,
       volume = {322},
       number = {5906},
        pages = {1348},
          doi = {10.1126/science.1166585},
archivePrefix = {arXiv},
       eprint = {0811.2606},
 primaryClass = {astro-ph},
       adsurl = {https://ui.adsabs.harvard.edu/abs/2008Sci...322.1348M}
}

@ARTICLE{macintosh2015,
       author = {{Macintosh}, B. and {Graham}, J.~R. and {Barman}, T. and {De Rosa}, R.~J. and {Konopacky}, Q. and {Marley}, M.~S. and {Marois}, C. and {Nielsen}, E.~L. and {Pueyo}, L. and {Rajan}, A. and {Rameau}, J. and {Saumon}, D. and {Wang}, J.~J. and {Patience}, J. and {Ammons}, M. and {Arriaga}, P. and {Artigau}, E. and {Beckwith}, S. and {Brewster}, J. and {Bruzzone}, S. and {Bulger}, J. and {Burningham}, B. and {Burrows}, A.~S. and {Chen}, C. and {Chiang}, E. and {Chilcote}, J.~K. and {Dawson}, R.~I. and {Dong}, R. and {Doyon}, R. and {Draper}, Z.~H. and {Duch{\^e}ne}, G. and {Esposito}, T.~M. and {Fabrycky}, D. and {Fitzgerald}, M.~P. and {Follette}, K.~B. and {Fortney}, J.~J. and {Gerard}, B. and {Goodsell}, S. and {Greenbaum}, A.~Z. and {Hibon}, P. and {Hinkley}, S. and {Cotten}, T.~H. and {Hung}, L. -W. and {Ingraham}, P. and {Johnson-Groh}, M. and {Kalas}, P. and {Lafreniere}, D. and {Larkin}, J.~E. and {Lee}, J. and {Line}, M. and {Long}, D. and {Maire}, J. and {Marchis}, F. and {Matthews}, B.~C. and {Max}, C.~E. and {Metchev}, S. and {Millar-Blanchaer}, M.~A. and {Mittal}, T. and {Morley}, C.~V. and {Morzinski}, K.~M. and {Murray-Clay}, R. and {Oppenheimer}, R. and {Palmer}, D.~W. and {Patel}, R. and {Perrin}, M.~D. and {Poyneer}, L.~A. and {Rafikov}, R.~R. and {Rantakyr{\"o}}, F.~T. and {Rice}, E.~L. and {Rojo}, P. and {Rudy}, A.~R. and {Ruffio}, J. -B. and {Ruiz}, M.~T. and {Sadakuni}, N. and {Saddlemyer}, L. and {Salama}, M. and {Savransky}, D. and {Schneider}, A.~C. and {Sivaramakrishnan}, A. and {Song}, I. and {Soummer}, R. and {Thomas}, S. and {Vasisht}, G. and {Wallace}, J.~K. and {Ward-Duong}, K. and {Wiktorowicz}, S.~J. and {Wolff}, S.~G. and {Zuckerman}, B.},
        title = "{Discovery and spectroscopy of the young jovian planet 51 Eri b with the Gemini Planet Imager}",
      journal = {Science},
         year = 2015,
        month = oct,
       volume = {350},
       number = {6256},
        pages = {64-67},
          doi = {10.1126/science.aac5891},
archivePrefix = {arXiv},
       eprint = {1508.03084},
 primaryClass = {astro-ph.EP},
       adsurl = {https://ui.adsabs.harvard.edu/abs/2015Sci...350...64M}
}

@ARTICLE{derosa2023,
       author = {{De Rosa}, Robert J. and {Nielsen}, Eric L. and {Wahhaj}, Zahed and {Ruffio}, Jean-Baptiste and {Kalas}, Paul G. and {Peck}, Anne E. and {Hirsch}, Lea A. and {Roberson}, William},
        title = "{Direct imaging discovery of a super-Jovian around the young Sun-like star AF Leporis}",
      journal = {\aap},
         year = 2023,
        month = apr,
       volume = {672},
          eid = {A94},
        pages = {A94},
          doi = {10.1051/0004-6361/202345877},
archivePrefix = {arXiv},
       eprint = {2302.06332},
 primaryClass = {astro-ph.EP},
       adsurl = {https://ui.adsabs.harvard.edu/abs/2023A&A...672A..94D}
}

@ARTICLE{franson2023,
       author = {{Franson}, Kyle and {Bowler}, Brendan P. and {Zhou}, Yifan and {Pearce}, Tim D. and {Bardalez Gagliuffi}, Daniella C. and {Biddle}, Lauren I. and {Brandt}, Timothy D. and {Crepp}, Justin R. and {Dupuy}, Trent J. and {Faherty}, Jacqueline and {Jensen-Clem}, Rebecca and {Morgan}, Marvin and {Sanghi}, Aniket and {Theissen}, Christopher A. and {Tran}, Quang H. and {Wolf}, Trevor N.},
        title = "{Astrometric Accelerations as Dynamical Beacons: A Giant Planet Imaged inside the Debris Disk of the Young Star AF Lep}",
      journal = {\apjl},
         year = 2023,
        month = jun,
       volume = {950},
       number = {2},
          eid = {L19},
        pages = {L19},
          doi = {10.3847/2041-8213/acd6f6},
archivePrefix = {arXiv},
       eprint = {2302.05420},
 primaryClass = {astro-ph.EP},
       adsurl = {https://ui.adsabs.harvard.edu/abs/2023ApJ...950L..19F}
}

@ARTICLE{mesa2023,
       author = {{Mesa}, D. and {Gratton}, R. and {Kervella}, P. and {Bonavita}, M. and {Desidera}, S. and {D'Orazi}, V. and {Marino}, S. and {Zurlo}, A. and {Rigliaco}, E.},
        title = "{AF Lep b: The lowest-mass planet detected by coupling astrometric and direct imaging data}",
      journal = {\aap},
         year = 2023,
        month = apr,
       volume = {672},
          eid = {A93},
        pages = {A93},
          doi = {10.1051/0004-6361/202345865},
archivePrefix = {arXiv},
       eprint = {2302.06213},
 primaryClass = {astro-ph.EP},
       adsurl = {https://ui.adsabs.harvard.edu/abs/2023A&A...672A..93M}
}

@ARTICLE{nielsen2019,
       author = {{Nielsen}, Eric L. and {De Rosa}, Robert J. and {Macintosh}, Bruce and {Wang}, Jason J. and {Ruffio}, Jean-Baptiste and {Chiang}, Eugene and {Marley}, Mark S. and {Saumon}, Didier and {Savransky}, Dmitry and {Ammons}, S. Mark and {Bailey}, Vanessa P. and {Barman}, Travis and {Blain}, C{\'e}lia and {Bulger}, Joanna and {Burrows}, Adam and {Chilcote}, Jeffrey and {Cotten}, Tara and {Czekala}, Ian and {Doyon}, Rene and {Duch{\^e}ne}, Gaspard and {Esposito}, Thomas M. and {Fabrycky}, Daniel and {Fitzgerald}, Michael P. and {Follette}, Katherine B. and {Fortney}, Jonathan J. and {Gerard}, Benjamin L. and {Goodsell}, Stephen J. and {Graham}, James R. and {Greenbaum}, Alexandra Z. and {Hibon}, Pascale and {Hinkley}, Sasha and {Hirsch}, Lea A. and {Hom}, Justin and {Hung}, Li-Wei and {Dawson}, Rebekah Ilene and {Ingraham}, Patrick and {Kalas}, Paul and {Konopacky}, Quinn and {Larkin}, James E. and {Lee}, Eve J. and {Lin}, Jonathan W. and {Maire}, J{\'e}r{\^o}me and {Marchis}, Franck and {Marois}, Christian and {Metchev}, Stanimir and {Millar-Blanchaer}, Maxwell A. and {Morzinski}, Katie M. and {Oppenheimer}, Rebecca and {Palmer}, David and {Patience}, Jennifer and {Perrin}, Marshall and {Poyneer}, Lisa and {Pueyo}, Laurent and {Rafikov}, Roman R. and {Rajan}, Abhijith and {Rameau}, Julien and {Rantakyr{\"o}}, Fredrik T. and {Ren}, Bin and {Schneider}, Adam C. and {Sivaramakrishnan}, Anand and {Song}, Inseok and {Soummer}, Remi and {Tallis}, Melisa and {Thomas}, Sandrine and {Ward-Duong}, Kimberly and {Wolff}, Schuyler},
        title = "{The Gemini Planet Imager Exoplanet Survey: Giant Planet and Brown Dwarf Demographics from 10 to 100 au}",
      journal = {\aj},
         year = 2019,
        month = jul,
       volume = {158},
       number = {1},
          eid = {13},
        pages = {13},
          doi = {10.3847/1538-3881/ab16e9},
archivePrefix = {arXiv},
       eprint = {1904.05358},
 primaryClass = {astro-ph.EP},
       adsurl = {https://ui.adsabs.harvard.edu/abs/2019AJ....158...13N}
}

@ARTICLE{astropy1,
       author = {{Astropy Collaboration} and {Robitaille}, Thomas P. and {Tollerud}, Erik J. and {Greenfield}, Perry and {Droettboom}, Michael and {Bray}, Erik and {Aldcroft}, Tom and {Davis}, Matt and {Ginsburg}, Adam and {Price-Whelan}, Adrian M. and {Kerzendorf}, Wolfgang E. and {Conley}, Alexander and {Crighton}, Neil and {Barbary}, Kyle and {Muna}, Demitri and {Ferguson}, Henry and {Grollier}, Fr{\'e}d{\'e}ric and {Parikh}, Madhura M. and {Nair}, Prasanth H. and {Unther}, Hans M. and {Deil}, Christoph and {Woillez}, Julien and {Conseil}, Simon and {Kramer}, Roban and {Turner}, James E.~H. and {Singer}, Leo and {Fox}, Ryan and {Weaver}, Benjamin A. and {Zabalza}, Victor and {Edwards}, Zachary I. and {Azalee Bostroem}, K. and {Burke}, D.~J. and {Casey}, Andrew R. and {Crawford}, Steven M. and {Dencheva}, Nadia and {Ely}, Justin and {Jenness}, Tim and {Labrie}, Kathleen and {Lim}, Pey Lian and {Pierfederici}, Francesco and {Pontzen}, Andrew and {Ptak}, Andy and {Refsdal}, Brian and {Servillat}, Mathieu and {Streicher}, Ole},
        title = "{Astropy: A community Python package for astronomy}",
      journal = {\aap},
         year = 2013,
        month = oct,
       volume = {558},
          eid = {A33},
        pages = {A33},
          doi = {10.1051/0004-6361/201322068},
archivePrefix = {arXiv},
       eprint = {1307.6212},
 primaryClass = {astro-ph.IM},
       adsurl = {https://ui.adsabs.harvard.edu/abs/2013A&A...558A..33A}
}

@ARTICLE{bowler2015,
       author = {{Bowler}, Brendan P. and {Liu}, Michael C. and {Shkolnik}, Evgenya L. and {Tamura}, Motohide},
        title = "{Planets around Low-mass Stars (PALMS). IV. The Outer Architecture of M Dwarf Planetary Systems}",
      journal = {\apjs},
         year = 2015,
        month = jan,
       volume = {216},
       number = {1},
          eid = {7},
        pages = {7},
          doi = {10.1088/0067-0049/216/1/7},
archivePrefix = {arXiv},
       eprint = {1411.3722},
 primaryClass = {astro-ph.EP},
       adsurl = {https://ui.adsabs.harvard.edu/abs/2015ApJS..216....7B}
}

@ARTICLE{biller2013,
       author = {{Biller}, Beth A. and {Liu}, Michael C. and {Wahhaj}, Zahed and {Nielsen}, Eric L. and {Hayward}, Thomas L. and {Males}, Jared R. and {Skemer}, Andrew and {Close}, Laird M. and {Chun}, Mark and {Ftaclas}, Christ and {Clarke}, Fraser and {Thatte}, Niranjan and {Shkolnik}, Evgenya L. and {Reid}, I. Neill and {Hartung}, Markus and {Boss}, Alan and {Lin}, Douglas and {Alencar}, Silvia H.~P. and {de Gouveia Dal Pino}, Elisabete and {Gregorio-Hetem}, Jane and {Toomey}, Douglas},
        title = "{The Gemini/NICI Planet-Finding Campaign: The Frequency of Planets around Young Moving Group Stars}",
      journal = {\apj},
         year = 2013,
        month = nov,
       volume = {777},
       number = {2},
          eid = {160},
        pages = {160},
          doi = {10.1088/0004-637X/777/2/160},
archivePrefix = {arXiv},
       eprint = {1309.1462},
 primaryClass = {astro-ph.EP},
       adsurl = {https://ui.adsabs.harvard.edu/abs/2013ApJ...777..160B}
}

@ARTICLE{greene1992,
       author = {{Greene}, Thomas P. and {Young}, Erick T.},
        title = "{Near-Infrared Observations of Young Stellar Objects in the rho Ophiuchi Dark Cloud}",
      journal = {\apj},
         year = 1992,
        month = aug,
       volume = {395},
        pages = {516},
          doi = {10.1086/171672},
       adsurl = {https://ui.adsabs.harvard.edu/abs/1992ApJ...395..516G}
}

@book{jeffreys1961theory,
  author = {Jeffreys, Harold},
  edition = {Third},
  publisher = {Oxford University Press},
  title = {The Theory of Probability},
  year = 1961
}
\bibliographystyle{aasjournal}



\end{document}